\documentclass{sig-alternate}

\usepackage{url}
\usepackage[breaklinks,bookmarks]{hyperref}
\usepackage{latexsym}
\usepackage{amsmath}
\usepackage{amssymb}
\usepackage{amsfonts}

\long\def\rem#1{}
\newcommand\comment[1]{}
\def\01{\{0,1\}}
\newcommand{\ket}[1]{|#1\rangle}
\newcommand{\bra}[1]{\langle #1|}
\newcommand{\inp}[2]{\langle #1|#2\rangle}
\newcommand{\lbra}{\langle}
\newcommand{\rket}{\rangle}

\newcommand{\ul}[1]{{\underline{#1}}}
\renewcommand{\Pr}{\mathrm{Pr}}
\newcommand{\Exp}{\mathrm{Exp}}
\newcommand{\Tr}{\mathop{\mathrm{Tr}}}
\newcommand{\vectorres}[2]{{#1^{(#2)}}}
\newcommand{\A}{\mathcal{A}}
\renewcommand{\H}{\mathcal{H}}
\renewcommand{\P}{\Pi}
\newcommand{\R}{\mathcal{R}}
\renewcommand{\S}{\mathcal{S}}
\newcommand{\BMP}{\textsc{Bounded Matrix Product}}
\newcommand{\SMP}{\textsc{Small Matrix Product}}

\newdef{definition}{Definition}
\newtheorem{theorem}{Theorem}
\newtheorem{lemma}[theorem]{Lemma}
\newtheorem{claim}[theorem]{Claim}
\newtheorem{corollary}[theorem]{Corollary}

\newcommand{\lemref}[1]{\hyperref[#1]{Lemma~\ref*{#1}}}
\newcommand{\thmref}[1]{\hyperref[#1]{Theorem~\ref*{#1}}}
\newcommand{\secref}[1]{\hyperref[#1]{Section~\ref*{#1}}}
\newcommand{\appref}[1]{\hyperref[#1]{Appendix~\ref*{#1}}}
\newcommand{\clmref}[1]{\hyperref[#1]{Claim~\ref*{#1}}}
\newcommand{\corref}[1]{\hyperref[#1]{Corollary~\ref*{#1}}}
\newcommand{\eqnref}[1]{\hyperref[#1]{(\ref*{#1})}}
\newcommand{\tabref}[1]{\hyperref[#1]{Table~\ref*{#1}}}

\begin{document}

\conferenceinfo{STOC'06,}{May 21--23, 2006, Seattle, Washington, USA.}
\CopyrightYear{2006}
\crdata{1-59593-134-1/06/0005}

\title{A New Quantum Lower Bound Method, with Applications to Direct Product
Theorems and Time-Space Tradeoffs}

\numberofauthors{3}
\author{
\alignauthor Andris Ambainis%
\titlenote{Institute for Quantum Computing and Department
of Combinatorics and Optimization, University of Waterloo.  Supported by
NSERC, ARO, CIAR and IQC University Professorship.}\\
\affaddr{University of Waterloo}\\
\email{ambainis@math.uwaterloo.ca}
\alignauthor Robert \v Spalek%
\titlenote{Supported in part by the European Commission under projects RESQ, IST-2001-37559, and QAP, IST-015848.}\\
\affaddr{CWI, Amsterdam}\\
\email{sr@cwi.nl}
\alignauthor Ronald de Wolf%
\titlenote{Supported by a Veni grant from the Netherlands Organization 
for Scientific Research (NWO) and partially supported by the EU projects RESQ and QAP.}\\
\affaddr{CWI, Amsterdam}\\
\email{rdewolf@cwi.nl}
}
\date{}
\maketitle

\begin{abstract}
We give a new version of the adversary method for proving lower
bounds on quantum query algorithms.  The new method is based on analyzing the
eigenspace structure of the problem at hand. We use it              
to prove a new and optimal strong direct product theorem for 2-sided error
quantum algorithms computing $k$ independent instances of a symmetric Boolean function:
if the algorithm uses significantly less than $k$ times the number of queries 
needed for one instance of the function, then its success probability is exponentially small in $k$.
We also use the polynomial method to prove a direct product theorem for 1-sided error 
algorithms for $k$ threshold functions with a stronger bound on the success probability.
Finally, we present a quantum algorithm for evaluating solutions to systems of linear
inequalities, and use our direct product theorems to show that the time-space
tradeoff of this algorithm is close to optimal.
\end{abstract}




\section{Introduction}

\subsection{A new adversary method}

Most of the known quantum algorithms work in the black-box model of computation. 
Here one accesses the $n$-bit input via \emph{queries} and our measure of complexity 
is the number of queries made by the algorithm.
In between the queries, the algorithm can make unitary transformations for free.
This model includes for instance the algorithms of Grover, Deutsch and Jozsa, Simon, quantum counting, the recent
quantum walk-based algorithms, and even Shor's period-finding algorithm (which is the quantum core of 
his factoring algorithm).

Much work has focused on proving \emph{lower bounds} in this model.
The two main methods known are the polynomial method and the adversary method.
The polynomial method \cite{nisan&szegedy:degree,bbcmw:polynomialsj}
works by lower-bounding the degree of a polynomial that in some way
represents the desired success probability. 

The adversary method was originally introduced by Ambainis~\cite{ambainis:lowerbounds}. 
Many different versions have since been
given~\cite{hns:search&sortj,bss:semidef,ambainis:degreevsquery,laplante&magniez:lower,zhang:ambainis},
but they are all equivalent~\cite{spalek&szegedy:adversaryj}.
Roughly speaking, the adversary method works as follows.
Suppose we have a $T$-query quantum algorithm that computes some function $f$ with high success probability.
Let $\ket{\psi_x^t}$ denote the algorithm's state on input $x$ after making the $t$-th query.
Suppose $x$ and $y$ are two inputs with distinct function values.
At the start of the algorithm ($t=0$), the states $\ket{\psi_x^0}$ and $\ket{\psi_y^0}$
are the same (the input has not been queried yet), so their inner product is $\inp{\psi_x^0}{\psi_y^0}=1$.  
But at the end of the algorithm ($t=T$), the inner product $\inp{\psi_x^T}{\psi_y^T}$
must be less than some small 
constant depending on the error probability, otherwise the algorithm cannot give the correct 
answer for both $x$ and $y$.  The adversary method takes a (weighted) sum of such
inner products (for $x,y$ pairs with $f(x)\neq f(y)$)
and analyzes how quickly this sum can go down after each new query.
If it cannot decrease quickly in one step, then it follows that we need many 
steps and we obtain a lower bound on $T$.

The two lower bound methods are incomparable. 
On the one hand, the adversary method proves stronger bounds than the polynomial
method for certain iterated functions~\cite{ambainis:degreevsquery},
and also gives tight lower bounds for constant-depth AND-OR
trees~\cite{ambainis:lowerbounds,hmw:berrorsearch},
where we do not know how to analyze the polynomial degree.
On the other hand, the polynomial method works well for analyzing zero-error 
or low-error quantum algorithms~\cite{bbcmw:polynomialsj,bcwz:qerror} and
gives optimal lower bounds for the collision problem and element
distinctness~\cite{aaronson&shi:collision}.
The adversary method fails for the latter problem (and also for other
problems like triangle-finding), because the best bound provable
with it is $O(\sqrt{C^0(f) C^1(f)})$~\cite{spalek&szegedy:adversaryj,zhang:ambainis}.
Here $C^0(f)$ and $C^1(f)$ are the certificate complexities of $f$ on 
0-inputs and 1-inputs.  In the case of element distinctness and triangle-finding, 
one of these complexities is constant. Hence the adversary method in its 
present form(s) can prove at most an $\Omega(\sqrt{N})$ bound,
while the true bound is $\Theta(N^{2/3})$~\cite{ambainis:ed} 
in the case of element distinctness and the best known algorithm for 
triangle-finding costs $O(N^{13/20})$~\cite{mss:triangle}.
A second limitation of the adversary method is that it cannot deal
well with the case where there are many different possible outputs,
and a success probability much smaller than $1/2$ would still be considered good.

In this paper we describe a new version of the adversary method that does
not suffer from the second limitation, and possibly also not from the first---though 
we have not found an example yet where the new method breaks through the
$\sqrt{C^0(f) C^1(f)}$ barrier.  

\emph{Very} roughly speaking, the new method works as follows.
We view the algorithm as acting on a 2-register state space ${\cal H}_A\otimes{\cal H}_I$.
Here the actual algorithm's operations take place in the first register, 
while the second contains (a superposition of) the inputs.
In particular, the query operation on ${\cal H}_A$ is now conditioned 
on the basis states in ${\cal H}_I$. We start the analysis with a superposition
of 0-inputs and 1-inputs in the input register, and then track how this register evolves
as the computation moves along. Let $\rho_t$ be the state of this register 
(tracing out the ${\cal H}_A$-register) after making the $t$-th query. 
By employing symmetries in the problem's structure, such as invariances of
the function under certain permutations of its input, we can decompose the input space into orthogonal
subspaces $S_0,\ldots,S_m$. We can decompose the state accordingly:
$$
\rho_t=\sum_{i=0}^m p_{t,i}\sigma_i,
$$
where $\sigma_i$ is a density matrix in subspace $S_i$.
Thus the $t$-th state can be fully described by a probability distribution 
$p_{t,0},\ldots,p_{t,m}$ that describes how the input register is 
distributed over the various subspaces.  Crucially, only some of the subspaces
are ``good'', meaning that the algorithm will only work if most of the
weight is concentrated in the good subspaces at the end of the computation.
At the start of the computation, hardly any weight will be in the good subspaces.
If we can show that in each query, not too much weight can move from the bad subspaces
to the good subspaces, then we again get a lower bound on $T$.

This idea was first introduced by Ambainis in~\cite{ambainis:dptor} and used there to reprove
the ``strong direct product theorem'' for the OR-function of~\cite{ksw:dpt} (we'll explain this in a minute).
In this paper we extend it and use it to prove direct product theorems for all \emph{symmetric} functions.

\subsection{Direct product theorems for symmetric\\ functions}

Consider an algorithm that simultaneously needs to compute
$k$ independent instances of a function $f$ (denoted $\vectorres{f}{k}$).
Direct product theorems deal with the optimal tradeoff between the resources
and success probability of such algorithms.
Suppose we need $t$ ``resources'' to compute a single instance $f(x)$ with
bounded error probability.  These resources could for example be time, space, ink,
queries, communication, etc.
A typical direct product theorem (DPT) has the following form:
\begin{quote}
Every algorithm with $T\leq\alpha k t$ resources for computing
$\vectorres{f}{k}$ has success probability $\sigma\leq 2^{-\Omega(k)}$
(where $\alpha>0$ is some small constant).
\end{quote}
This expresses our intuition that essentially the best way to compute
$\vectorres{f}{k}$ on $k$ independent instances is to run separate
$t$-resource algorithms for each of the instances.
Since each of those will have success probability less than 1, we expect that
the probability of simultaneously getting all $k$ instances right
goes down exponentially with $k$.
DPT's can be stated for classical algorithms or quantum algorithms,
and $\sigma$ could measure worst-case success probability or average-case
success probability under some input distribution.
DPT's are generally hard to prove, and Shaltiel~\cite{shaltiel:sdpt}
even gives general examples where they are just not true
(with $\sigma$ average success probability), the above intuition
notwithstanding. 
Klauck, \v{S}palek, and de~Wolf~\cite{ksw:dpt} recently examined
the case where the resource is query complexity and $f=$ OR, and proved
an optimal DPT both for classical algorithms and for quantum algorithms
(with $\sigma$ worst-case success probability).
This strengthened a slightly earlier result of Aaronson~\cite{aaronson:advicecomm},
who proved that the success probability goes down exponentially with $k$
if the number of queries is bounded by $\alpha\sqrt{kn}$ rather than the $\alpha k\sqrt{n}$ of~\cite{ksw:dpt}.

Here we generalize their results to
the case where $f$ can be any symmetric function, i.e., a function
depending only on the Hamming weight $|x|$ of its input.
In the case of classical algorithms the situation is quite simple.
Every $n$-bit symmetric function $f$ has classical bounded-error query complexity
$R_2(f)=\Theta(n)$ and block sensitivity $bs(f)=\Theta(n)$, hence
an optimal classical DPT follows immediately from~\cite[Theorem~3]{ksw:dpt}.
Classically, all symmetric functions essentially ``cost the same'' in terms
of query complexity. This is different in the quantum world.
For instance, the OR function has bounded-error quantum query complexity
$Q_2(\mathrm{OR})=\Theta(\sqrt{n})$~\cite{grover:search,bhmt:countingj},
while Parity needs $n/2$ quantum queries~\cite{bbcmw:polynomialsj,fggs:parity}.
If $f$ is a $t$-threshold function ($f(x)=1$ iff $|x|\geq t$, with $t\leq n/2$), 
then $Q_2(f)=\Theta(\sqrt{tn})$~\cite{bbcmw:polynomialsj}.

Our main result is an essentially optimal quantum DPT for all symmetric functions:
\begin{quote}
There is a constant $\alpha>0$ such that for every symmetric
$f$ and every positive integer $k$:
Every 2-sided error quantum algorithm with $T\leq\alpha k Q_2(f)$ queries
for computing $\vectorres{f}{k}$ has success probability $\sigma\leq 2^{-\Omega(k)}$.
\end{quote}
Our new direct product theorem generalizes the polynomial-based results of~\cite{ksw:dpt}
(which strengthened the polynomial-based~\cite{aaronson:advicecomm}), but our
current proof uses the above-mentioned version of the adversary method.

We have not been able to prove this result using the polynomial method. 
We can, however, use the polynomial method to prove an incomparable DPT.
This result is worse than our main result in applying only to 
\emph{1-sided error} quantum algorithms%
\footnote{The error is 1-sided if 1-bits in the $k$-bit output vector are always correct.} 
for \emph{threshold} functions; but it's better in
giving a much stronger upper bound on the success probability:
\begin{quote}
There is a constant $\alpha>0$ such that for every $t$-threshold function $f$ and every positive integer $k$:
Every 1-sided error quantum algorithm with $T\leq\alpha k Q_2(f)$ queries
for computing $\vectorres{f}{k}$ has success probability $\sigma\leq 2^{-\Omega(kt)}$.
\end{quote}
A similar theorem can be proven for the $k$-fold $t$-search
problem, where in each of $k$ inputs of $n$ bits, we want to find at least $t$ ones.
The different error bounds $2^{-\Omega(kt)}$ and $2^{-\Omega(k)}$ for 
1-sided and 2-sided error algorithms intuitively say that imposing the 1-sided
error constraint makes deciding each of the $k$ threshold problems 
as hard as actually \emph{finding} $t$ ones in each of the $k$ inputs.

\subsection{Time-Space tradeoffs for evaluating solutions to systems of linear inequalities}

As an application we obtain near-optimal time-space tradeoffs
for evaluating solutions to systems of linear equalities. Such tradeoffs between the two
main computational resources are well known classically for problems
like sorting, element distinctness, hashing, etc.
In the quantum world, essentially optimal time-space tradeoffs were recently obtained
for sorting and for Boolean matrix multiplication~\cite{ksw:dpt}, but little else is known.

Let $A$ be a fixed $N\times N$ matrix of nonnegative integers.
Our inputs are column vectors $x=(x_1,\ldots,x_N)$ and $b=(b_1,\ldots,b_N)$
of nonnegative integers. We are interested in the system
$$
Ax\geq b
$$
of $N$ linear inequalities, and want to find out which of these inequalities hold
(we could also mix $\geq$, $=$, and $\leq$, but omit that for ease of notation).%
\footnote{Note that if $A$ and $x$ are Boolean and $b=(t,\ldots,t)$, this
gives $N$ overlapping $t$-threshold functions.}
Note that the output is an $N$-bit vector.
We want to analyze the tradeoff between the time $T$ and space $S$ needed
to solve this problem. Lower bounds on $T$ will be in terms of query complexity.
For simplicity we omit polylog factors in the following discussion.

In the classical world, the optimal tradeoff is $TS=N^2$, independent of
the values in $b$. This follows from~\cite[Section~7]{ksw:dpt}.
The upper bounds are for deterministic algorithms and the lower bounds
are for 2-sided error algorithms.
In the quantum world the situation is more complex. Let us put an upper bound $\max\{b_i\}\leq t$.
We have two regimes for 2-sided error quantum algorithms:
\begin{itemize}
\item Quantum regime. If $S\leq N/t$ then the optimal tradeoff is $T^2 S=t N^3$
(better than classical).
\item Classical regime. If $S>N/t$ then the optimal tradeoff is $TS=N^2$
(same as classical).
\end{itemize}
Our lower bounds hold even 
for the constrained situation where $b$ is fixed to the all-$t$ vector, 
$A$ and $x$ are Boolean, and $A$ is sparse in having only 
$O(N/S)$ non-zero entries in each row.

Since our DPT for 1-sided error algorithms is stronger by
an extra factor of $t$ in the exponent, we obtain a stronger lower bound
for 1-sided error algorithms:
\begin{itemize}
\item If $t\leq S\leq N/t^2$ then the optimal tradeoff for 1-sided error algorithms is $T^2S\geq t^2 N^3$.
\item If $S>N/t^2$ then the optimal tradeoff for 1-sided error algorithms is $TS=N^2$.
\end{itemize}
We do not know whether the lower bound in the first case is optimal (probably it is not),
but note that it is stronger than the optimal bounds that we have for 2-sided error algorithms.
This is the first separation of 2-sided and 1-sided error algorithms 
in the context of quantum time-space tradeoffs.\footnote{Strictly speaking,
there's a quadratic gap for OR, but space $\log n$ suffices for the fastest 1-sided 
and 2-sided error algorithms so there's no real tradeoff in that case.}

\medskip

\noindent
{\bf Remarks:}

1. Klauck et al.~\cite{ksw:dpt} gave direct product theorems not only for quantum query
complexity, but also for 2-party quantum communication complexity, and derived some
communication-space tradeoffs in analogy to the time-space tradeoffs.
This was made possible by a translation of communication protocols to polynomials
due to Razborov~\cite{razborov:qdisj}, and the fact that the DPTs of~\cite{ksw:dpt} were
polynomial-based.  Some of the results in this paper can similarly be ported to
a communication setting, though only the ones that use the polynomial method.

2. The time-space tradeoffs for 2-sided error algorithms for $Ax\geq b$ similarly
hold for a system of $N$ equalities, $Ax=b$. The upper bound clearly carries over,
while the lower holds for equalities as well, because our DPT holds even under the
promise that the input has weight $t$ or $t-1$.
In contrast, the stronger 1-sided error time-space tradeoff does not automatically carry over
to systems of equalities, because we do not know how to prove the DPT with bound $2^{-\Omega(kt)}$
under this promise.

\section{Preliminaries}

We assume familiarity with quantum computing~\cite{nielsen&chuang:qc}
and sketch the model of quantum query complexity,
referring to~\cite{buhrman&wolf:dectreesurvey} for more
details, also on the close relation between query complexity
and degrees of multivariate polynomials.
Suppose we want to compute some function $f$.
For input $x\in\01^N$, a \emph{query} gives us access to the
input bits. It corresponds to the unitary transformation
$$
O_x:\ket{i,b,z}\mapsto\ket{i,b\oplus x_i,z}.
$$
Here $i\in[N]=\{1,\ldots,N\}$ and $b\in\01$; the $z$-part corresponds
to the workspace, which is not affected by the query.
We assume the input can be accessed only via such queries.
A $T$-query quantum algorithm has the form $A=U_T O_x U_{T-1}\cdots O_x U_1 O_x U_0$,
where the $U_k$ are fixed unitary transformations, independent of $x$.
This $A$ depends on $x$ via the $T$ applications of $O_x$.
The algorithm starts in initial $S$-qubit state $\ket{0}$ and
its \emph{output} is the result of measuring a dedicated part
of the final state $A\ket{0}$.
For a Boolean function $f$, the output
of $A$ is obtained by observing the leftmost qubit of the
final superposition $A\ket{0}$, and its \emph{acceptance probability} on input $x$
is its probability of outputting 1.
We mention some well known quantum algorithms that we use as subroutines.
\begin{itemize}
\item {\bf Quantum search.}
Grover's search algorithm~\cite{grover:search,bbht:bounds}
can find an index of a 1-bit in an $n$-bit input in
expected number of $O(\sqrt{n/(|x|+1)})$ queries,
where $|x|$ is the Hamming weight (number of ones) in the input.
If $|x|$ is known, the algorithm can be made to find
the index in exactly $O(\sqrt{n/(|x|+1)})$ queries,
instead of the expected number~\cite{bhmt:countingj}.
By repeated search, we can find $t$ ones in an $n$-bit input
with $|x|\geq t$, using
$\sum_{i=|x|-t+1}^{|x|} O(\sqrt{n/(i+1)})=O(\sqrt{tn})$ queries.
\item {\bf Quantum counting}~\cite[Theorem 13]{bhmt:countingj}.
There is a quantum algorithm that uses $M$ queries to $n$-bit $x$ to
compute an estimate $w$ of $|x|$ such that with probability at least $8 / \pi^2$
\[
|w - |x|| \le 2 \pi {\sqrt{|x| (n-|x|)} \over M} + \pi^2 \frac n {M^2}.
\]
\end{itemize}
For investigating time-space tradeoffs we use the circuit model. A
circuit accesses its input via an oracle like a query algorithm.
Time corresponds to the number of gates in the circuit. We will,
however, usually consider the number of queries to the input,
which is obviously a lower bound on time. A circuit uses
space $S$ if it works with $S$ bits/qubits only. We require that the
outputs are made at predefined gates in the circuit, by writing
their value to some extra bits/qubits that may not be used later on.

\section{Direct Product Theorem for\\ Symmetric Functions (2-sided)}\label{secdptsym}

The main result of this paper is the following theorem.

\begin{theorem}\label{thsdpt2sided}
There is a constant $\alpha>0$ such that for every symmetric
$f$ and every positive integer $k$:
Every 2-sided error quantum algorithm with $T\leq\alpha k Q_2(f)$ queries
for computing $\vectorres{f}{k}$ has success probability $\sigma\leq 2^{-\Omega(k)}$.
\end{theorem}

Let us first say something about $Q_2(f)$ for a symmetric function $f:\01^n\rightarrow\01$.
Let $t$ denote the smallest nonnegative integer such that $f$ is constant on the interval $|x|\in[t,n-t]$. 
We call this value $t$ the ``implicit threshold'' of $f$.  For instance, 
functions like OR and AND have $t=1$, 
while Parity and Majority have $t=n/2$. If $f$ is the $t$-threshold function, 
then the implicit threshold is just the threshold.
The implicit threshold is related to the parameter $\Gamma(f)$ introduced by 
Paturi~\cite{paturi:degree} via $t=n/2-\Gamma(f)/2\pm 1$.
It characterizes the bounded-error quantum query complexity of $f$:  
$Q_2(f)=\Theta(\sqrt{tn})$~\cite{bbcmw:polynomialsj}.
Hence our resource bound in the above theorem will be $\alpha k \sqrt{tn}$ for some small constant $\alpha>0$.

We actually prove a stronger statement, applying to any Boolean function $f$
(total or partial) for which $f(x)=0$ if $|x|=t-1$ and $f(x)=1$ if $|x|=t$.
In this section we give an outline of the proof. 
Most of the proofs of technical claims are deferred to
\appref{app:andrisproof}.

Let $\A$ be an algorithm that computes $k$ instances of this 
weight-$(t-1)$ versus weight-$t$ problem. 
\comment{We first ``symmetrize" $\A$ by adding an extra register $\H_P$ 
holding $k$ permutations $\pi_1, \ldots, \pi_k\in S_n$. 

Initially, $\H_P$ holds a uniform superposition
of all $k$-tuples of permutations $(\pi_1, \ldots, \pi_k)$: 
$\frac{1}{(n!)^{k/2}}\sum_{\pi_1, \ldots, \pi_k\in S_n} \ket{\pi_1, \ldots, \pi_k}$. 
Before each query $O$, we insert a transformation 
\[ \ket{j, i}\ket{\pi_1, \ldots, \pi_k}
\rightarrow\ket{j, \pi_j^{-1}(i)}
\ket{\pi_1, \ldots, \pi_k} \]
on the part of the state containing the index $(j, i)$ 
to be queried and $\H_P$. 
After the query, we insert a transformation
\[ \ket{j, i}\ket{\pi_1, \ldots, \pi_k}
\rightarrow\ket{j, \pi_j(i)}
\ket{\pi_1, \ldots, \pi_k} .\]
At the end of algorithm, we apply the transformation
\[ \ket{i_1} \ldots \ket{i_k}\ket{\pi_1, \ldots, \pi_k}\rightarrow 
\ket{\pi^{-1}_1(i_1)}\ldots\ket{\pi^{-1}_k(i_k)}\ket{\pi_1, \ldots, 
\pi_{k}}. \] 
The effect of the symmetrization 
is that, on the subspace $\ket{x^1, \ldots, x^k}\otimes 
\ket{\pi_1, \ldots, \pi_k}$, 
the algorithm is effectively running on the input 
$\pi_1(x^1)$, $\ldots$, $\pi_k(x^k)$ (where
$\pi_i(x^i)$ is defined by $(\pi_i(x^i))_j = x^i_{\pi_i(j)}$). 
If the original algorithm $\A$ succeeds on every input $(x^1, \ldots, x^k)$
with probability at least $\epsilon$, the symmetrized algorithm
also succeeds with probability at 
least $\epsilon$, since its success probability is just 
the average of the success probabilities of $\A$ over all $(x^1, \ldots, x^k)$ 
with the same $|x^1|, \ldots, |x^k|$.

Next,}
We recast $\A$ into a different form, 
using a register that stores the input $x^1, \ldots, x^k$.
Let $\H_A$ be the Hilbert space on which 
$\A$ operates. 
Let $\H_I$ be an $({n \choose t-1}+{n \choose t})^k$-dimensional 
Hilbert space whose basis states correspond to inputs 
$(x^1, \ldots, x^k)$ with Hamming weights $|x^1|\in\{t-1, t\},\ldots,|x^k|\in\{t-1, t\}$. 
We transform $\A$ into a sequence of transformations on 
a Hilbert space $\H=\H_A\otimes\H_I$. 
A non-query transformation $U$ on $\H_A$ is replaced with 
$U\otimes I$ on $\H$. 
A query is replaced by a transformation $O$ that is equal 
to $O_{x^1, \ldots, x^k}\otimes I$
on the subspace consisting of states of the form 
$\ket{s}_A\otimes\ket{x^1 \ldots x^k}_I$.
The starting state of the algorithm on Hilbert space $\H$ is 
$\ket{\varphi_0}=\ket{\psi_{start}}_A\otimes\ket{\psi_0}_I$ 
where $\ket{\psi_{start}}$ is the starting state of $\A$ 
as an algorithm acting on $\H_A$ and 
$\ket{\psi_0}=\ket{\psi_{one}}^{\otimes k}$
is a tensor product of $k$ copies of the state $\ket{\psi_{one}}$
in which half of the weight is on $\ket{x}$ with $|x|=t$,
the other half is on $\ket{x}$ with $|x|=t-1$, and any two states 
$\ket{x}$ with the same $|x|$ have equal amplitudes: 
\[ \ket{\psi_{one}}=\frac{1}{\sqrt{2{n \choose t}}}
\sum_{x:|x| = t} \ket{x} +\frac{1}{\sqrt{2{n \choose t-1}}}
\sum_{x:|x| = t-1} \ket{x} .\]
Let $\ket{\varphi_d}$ be the state of the 
algorithm $\A$, as a sequence of transformations 
on $\H$, after the $d$-th query. 
Let $\rho_d$ be the mixed state in $\H_I$ obtained from
$\ket{\varphi_d}$ by tracing out the $\H_A$ register. 

We define two decompositions of $\H_I$ into a direct sum of subspaces.
We have $\H_I=(\H_{one})^{\otimes k}$ where $\H_{one}$ is the input Hilbert
space for one instance, with basis states $\ket{x}$, $x\in\{0, 1\}^n, |x|\in\{t-1, t\}$.
Let
\[ \ket{\psi^0_{i_1, \ldots, i_j}}=\frac{1}{\sqrt{n-j \choose t-1-j}}\mathop{\mathop{\sum_{x_1, \ldots, x_n:}}_{x_1+\cdots+x_n=t-1,}}_{x_{i_1}=\cdots=x_{i_j}=1} 
\ket{x_1\ldots x_n} \]
and let $\ket{\psi^1_{i_1, \ldots, i_j}}$ be a similar  
state with $x_1+\cdots+x_n=t$ instead of $x_1+\cdots+x_n=t-1$.
Let $T_{j,0}$ (resp.~$T_{j, 1}$) be the space spanned by all states
$\ket{\psi^0_{i_1, \ldots, i_j}}$ (resp.~$\ket{\psi^{1}_{i_1, \ldots, i_j}}$) 
and let $S_{j, a}=T_{j, a}\cap T_{j-1, a}^\perp$.
For a subspace $S$, we use $\P_S$ to denote the projector onto $S$.
Let $\ket{\tilde{\psi}^a_{i_1, \ldots, i_j}}=
\Pi_{T_{j-1, a}^\perp}\ket{\psi^a_{i_1, \ldots, i_j}}$.
For $j<t$, let $S_{j, +}$ be the subspace spanned by the states
\[ \frac{\ket{\tilde{\psi}^0_{i_1, \ldots, i_j}}}{
\|\tilde{\psi}^0_{i_1, \ldots, i_j}\|} + 
\frac{\ket{\tilde{\psi}^1_{i_1, \ldots, i_j}}}{
\|\tilde{\psi}^1_{i_1, \ldots, i_j}\|} \]
and $S_{j, -}$ be the subspace spanned by 
\[ \frac{\ket{\tilde{\psi}^0_{i_1, \ldots, i_j}}}{
\|\tilde{\psi}^0_{i_1, \ldots, i_j}\|} -
\frac{\ket{\tilde{\psi}^1_{i_1, \ldots, i_j}}}{
\|\tilde{\psi}^1_{i_1, \ldots, i_j}\|} \]
For $j=t$, we define $S_{t, -}=S_{t, 1}$
and there is no subspace $S_{t, +}$.
Thus $\H_{one}=\bigoplus_{j=0}^{t-1} (S_{j, +}\oplus S_{j, -}) \oplus S_{t, -}$.
Let us try to give some intuition.  In the spaces $S_{j,+}$ and $S_{j,-}$, 
we may be said to ``know'' the positions of $j$ of the ones.  In the $S_{j,-}$ subspaces
we have distinguished the 0-inputs from 1-inputs by the relative phase, while
in the $S_{j,+}$ subspace we have not distinguished them.  Accordingly,
the algorithm is doing well on this one instance if most of the state 
sits in the ``good'' subspaces $S_{j,-}$.

For the space $\H_I$ (representing $k$ independent inputs for our function)
and $r_1, \ldots, r_k\in\{+,-\}$,
we define
\[ 
S_{j_1, \ldots, j_k, r_1, \ldots, r_k} =
S_{j_1, r_1} \otimes S_{j_2, r_2} \otimes \cdots
\otimes S_{j_k, r_k} .
\]
Let $\S_{m-}$ be the direct sum of all $S_{j_1,\ldots,j_k, r_1,\dots,r_k}$
such that exactly $m$ of the signs $r_1, \ldots, r_k$ are equal to $-$.
Then $\H_I=\bigoplus_m \S_{m-}$.  This is the first decomposition.

The above intuition for one instance carries over to $k$ instances:
the more minuses the better for the algorithm. Conversely, if most
of the input register sits in $S_{m-}$ for low $m$, then its success
probability will be small.
More precisely, in \appref{app:proof1} we prove:

\begin{lemma}
\label{lem:prob}
Let $\rho$ be the reduced density matrix of $\H_I$.
If the support of $\rho$ is contained in 
$\S_{0-}\oplus \S_{1-}\oplus\cdots \oplus \S_{m-}$,
then the probability that measuring $\H_A$ gives the correct answer
is at most $\displaystyle\frac{\sum_{m'=0}^m {k \choose m'}}{2^k}$.
\end{lemma}

Note that this probability is exponentially small in $k$ for, say, $m=k/3$.
The following consequence of this lemma is proven in 
\appref{app:proof1cor}:

\begin{corollary}
\label{cor:prob}
Let $\rho$ be the reduced density matrix of $\H_I$.
The probability that measuring $\H_A$ gives the correct answer
is at most 
\[ \frac{\sum_{m'=0}^m {k \choose m'}}{2^k} + 4
\sqrt{\Tr \P_{(\S_{0-}\oplus \S_{1-}\oplus\cdots \oplus \S_{m-})^{\perp}} \rho} .\] 
\end{corollary}

To define the second decomposition, we express
$\H_{one}=\bigoplus_{j=0}^{t/2} R_j$ with $R_j=S_{j, +}$ for $j<t/2$
and 
\[ R_{t/2}=\bigoplus_{j\geq t/2} S_{j, +} \oplus \bigoplus_{j\geq 0} S_{j, -}. \]
Intuitively, all subspaces except for $R_{t/2}$ are ``bad'' for the algorithm,
since they equal the ``bad'' $S_{j,+}$ subspaces.
Let $\R_\ell$ be the direct sum of all $R_{j_1}\otimes  
\cdots \otimes R_{j_k}$ satisfying $j_1+\cdots+j_k=\ell$.
Then $\H_I=\bigoplus_{\ell=0}^{t k/2} \R_\ell$.
This is the second decomposition.

Intuitively, the algorithm can only have good success probability if
for most of the $k$ instances, most of the input register sits in $R_{t/2}$.
Aggregated over all $k$ instances, this means that the algorithm will only
work well if most of the $k$-input register sits in $\R_\ell$ for $\ell$ large, 
meaning fairly close to $kt/2$.
Our goal below is to show that this cannot happen if the number of queries is small.

Let $\R'_j=\bigoplus_{\ell=j}^{t k/2} \R_\ell$.
Note that $\S_{m-}\subseteq \R'_{t m/2}$ for every $m$: 
$\S_{m-}$ is the direct sum of subspaces 
$S=S_{j_1,r_1}\otimes\cdots\otimes S_{j_k,r_k}$ having
$m$ minuses among $r_1,\ldots,r_k$; each such minus-subspace sits in the corresponding
$R_{t/2}$ and hence $S\subseteq\R'_{t m/2}$.
This implies
\[
(\S_{0-}\oplus \S_{1-}\oplus\cdots \oplus \S_{(m-1)-})^{\perp}\subseteq \R'_{t m/2}.
\]
Accordingly, if we prove an upper bound on $\Tr \P_{\R'_{t m/2}} \rho_{T}$,
where $T$ is the total number of queries,
this bound together with \corref{cor:prob} implies
an upper bound on the success probability of $\A$.
To bound $\Tr \P_{\R'_{t m/2}} \rho_{T}$, we consider the following potential function
\[ P(\rho) = \sum_{m=0}^{t k/2} q^m \Tr \P_{\R_{m}} \rho ,\]
where $q=1+\frac{1}{t}$.
Then for every $d$
\begin{equation}
\label{eq:potential}
 \Tr \P_{\R'_{t m/2}} \rho_{d} \leq P(\rho_d) q^{-t m/2}
= P(\rho_d) e^{-(1+o(1))m/2}. 
\end{equation}

$P(\rho_0)=1$, because the initial state $\ket{\psi_0}$ is a tensor
product of the states $\ket{\psi_{one}}$ 
on each copy of $\H_{one}$ and $\ket{\psi_{one}}$ belongs
to $S_{0, +}$, hence $\ket{\psi_0}$ belongs to $\R_0$. 
In~\appref{app:proof2} we prove

\begin{lemma}
\label{lem:onestep}
There is a constant $C$ such that
\[
P(\rho_{j+1}) \leq  \left(1+ \frac{C}{\sqrt{t n}} (q^{t/2}-1) + 
\frac{C\sqrt{t}}{\sqrt{n}} (q-1) \right)P(\rho_j).
\]
\end{lemma}

Since $q=1+\frac{1}{t}$, \lemref{lem:onestep} means that
$P(\rho_{j+1})\leq (1+\frac{C\sqrt{e}}{\sqrt{t n}}) P(\rho_{j})$ and
$P(\rho_{j}) \leq (1+\frac{C\sqrt{e}}{\sqrt{t n}})^j\leq 
e^{2 C j/\sqrt{t n}}$. By equation \eqnref{eq:potential}, for the final state after $T$ queries we have
\[ 
\Tr \P_{\R'_{t m/2}} \rho_{T} \leq e^{2 C T /\sqrt{t n}-(1+o(1))m/2} .
\]
We take $m=k/3$.
Then if $T\leq m \sqrt{t n}/8C$, 
this expression is exponentially small in $k$. Together with
\corref{cor:prob}, this implies the theorem.

\section{Direct Product Theorem for\\ Threshold Functions (1-sided)}\label{secdptthresh}

The previous section used the adversary method to prove a direct product theorem
for 2-sided error algorithms computing $k$ instances of some symmetric function.
In this section we use the polynomial method to obtain stronger direct product theorems
for 1-sided error algorithms for threshold functions. An algorithm for $\vectorres{f}{k}$
has 1-sided error if the 1's in its $k$-bit output vector are always correct.

Our use of polynomials is a relatively small extension of the argument in~\cite{ksw:dpt}.
We use three results about polynomials, also used in~\cite{bcwz:qerror,ksw:dpt}.
The first is by Coppersmith and Rivlin~\cite[p.~980]{coppersmith&rivlin:poly}
and gives a general bound for polynomials bounded by 1 at integer points:

\begin{theorem}
[Coppersmith \&~Rivlin~\cite{coppersmith&rivlin:poly}]
\label{thcopprivlin}
Every polynomial $p$ of degree $d \le n$ that has absolute value
\[
|p(i)|\leq 1 \mbox{ for all integers } i\in[0,n],
\]
satisfies
\[
|p(x)|< a e^{b d^2/n} \mbox{ for all real } x\in[0,n],
\]
where $a,b>0$ are universal constants (no explicit values for $a$ and $b$
are given in~\cite{coppersmith&rivlin:poly}).
\end{theorem}

The other two results concern the Chebyshev polynomials $T_d$, defined
as in~\cite{rivlin:chebyshev}:
\[
T_d(x)=\frac{1}{2}
\left(\left(x+\sqrt{x^2-1}\right)^d+\left(x-\sqrt{x^2-1}\right)^d\right).
\]
$T_d$ has degree $d$ and its absolute value $|T_d(x)|$ is bounded by 1 if
$x\in[-1,1]$.  On the interval $[1,\infty)$, $T_d$ exceeds all others
polynomials with those two properties (\cite[p.108]{rivlin:chebyshev}
and~\cite[Fact~2]{paturi:degree}):

\begin{theorem}
\label{thchebextremal}
If $q$ is a polynomial of degree $d$ such that $|q(x)|\leq 1$ for all
$x\in[-1,1]$ then $|q(x)|\leq|T_d(x)|$ for all $x\geq 1$.
\end{theorem}


\begin{lemma}
[Paturi~\cite{paturi:degree}]
\label{lemchebbound}
$T_d(1+\mu)\leq e^{2d\sqrt{2\mu+\mu^2}}$ for all $\mu\geq 0$.
\end{lemma}

\begin{proof}
For $x=1+\mu$:
$\displaystyle T_d(x)\leq (x+\sqrt{x^2-1})^d=(1+\mu+\sqrt{2\mu+\mu^2})^d\leq
(1+2\sqrt{2\mu+\mu^2})^d \leq e^{2d\sqrt{2\mu+\mu^2}}$.
\end{proof}

The following lemma is key. It analyzes polynomials that are 0
on the first $m$ integer points, and that significantly ``jump'' a bit later.

\begin{lemma}\label{lem:keypol}
Suppose $E,N,m$ are integers satisfying $10 \le E \le \frac N {2 m}$,
and let $p$ be a degree-$D$ polynomial such that
\begin{quote}
$p(i)=0$ for all $i\in\{0,\ldots,m-1\}$,\\[1mm]
$p(8m)=\sigma$,\\[1mm]
$p(i)\in[0,1]$ for all $i\in\{0,\ldots,N\}$.
\end{quote}
Then $\displaystyle\sigma\leq 2^{O(D^2/N+D\sqrt{E m/N}-m \log E)}$.
\end{lemma}

\begin{proof}
Divide $p$ by $\prod_{j=0}^{m-1} (x-j)$ to obtain
\[
p(x) = q(x) \prod_{j=0}^{m-1} (x-j),
\]
where $d=\deg(q)=D-m$.
This implies the following about the values of
the polynomial $q$:
\begin{eqnarray*}
|q(8m)| &\geq& \sigma/(8m)^m, \\
|q(i)|  &\leq& 1/((E-1) m)^m
\quad \mbox{for $i\in\{E m,\ldots,N\}$}. \\
\noalign{\noindent \thmref{thcopprivlin} implies that there are constants
$a,b>0$ such that \medskip}
|q(x)| &\leq& \frac{a}{((E-1)m)^m}e^{b d^2/(N-E m)}=B \\
&& \mbox{for all real $x\in[E m,N]$}.
\end{eqnarray*}
We now divide $q$ by $B$ to normalize it, and
rescale the interval $[E m,N]$ to $[1,-1]$ to get a degree-$d$ polynomial $t$
satisfying
\begin{eqnarray*}
|t(x)| &\leq& 1 \quad \mbox{for all $x\in[-1,1]$}, \\
t(1+\mu) &=& q(8m)/B \quad \mbox{for $\mu=2(E-8) m/(N-E m)$}. \\
\noalign{\medskip\noindent Since $t$ cannot grow faster than the degree-$d$
Chebyshev polynomial, \thmref{thchebextremal} and \lemref{lemchebbound} imply\medskip} 
t(1+\mu) &\leq& e^{2d\sqrt{2\mu+\mu^2}}.
\end{eqnarray*}
Combining our upper and lower bounds on $t(1+\mu)$ gives
\[
\frac{\sigma}{(8m)^m}\cdot\frac{((E-1) m)^m}{a e^{O(d^2/N)}}
\leq e^{O(d\sqrt{E m/N})},
\]
which implies the lemma.
\end{proof}

\begin{theorem}\label{thsdpt1sided}
There exists $\alpha>0$ such that for every threshold function $T_t$ and positive integer $k$:
Every 1-sided error quantum algorithm with $T\leq\alpha k Q_2(T_t)$ queries
for computing $\vectorres{T_t}{k}$ has success probability $\sigma\leq 2^{-\Omega(kt)}$.
\end{theorem}

\begin{proof}
We assume without loss of generality that $t\leq n/20$,
the other cases can easily be reduced to this.
We know that $Q_2(T_t)=\Theta(\sqrt{tn})$~\cite{bbcmw:polynomialsj}.
Consider a quantum algorithm $A$ with $T\leq\alpha k\sqrt{tn}$ queries
that computes $\vectorres{f}{k}$ with success probability $\sigma$.
Roughly speaking, we use $A$ to solve one big threshold problem on
the total input, and then invoke the polynomial lemma to upper
bound the success probability.

Define a new quantum algorithm $B$ on an input $x$ of $N=kn$ bits,
as follows: $B$ runs $A$ on a random permutation $\pi(x)$,
and then outputs 1 iff the $k$-bit output vector has at least $k/2$ ones.

Let $m=kt/2$.
Note that if $|x|<m$, then $B$ always outputs 0 because the 1-sided
error output vector must have fewer than $k/2$ ones.
Now suppose $|x|=8m=4kt$. Call an $n$-bit input block ``full'' if
$\pi(x)$ contains at least $t$ ones in that block. Let $F$ be
the random variable counting how many of the $k$ blocks are full.
We claim that $\Pr[F\geq k/2]\geq 1/9$.
To prove this, observe that the number $B$ of ones in one fixed block
is a random variable distributed according to a hypergeometric distribution
($4kt$ balls into $N$ boxes, $n$ of which count as success) with expectation
$\mu=4 t$ and variance $V\leq 4 t$. Using Chebyshev's inequality
we bound the probability that this block is not full:
\begin{align*}
\Pr[B<t] &\leq \Pr[|B-\mu|>3 t]\leq \Pr[|B-\mu|>(3\sqrt{t}/2)\sqrt{V}]\\
         &<\frac{1}{(3\sqrt{t}/2)^2}\leq \frac{4}{9}.
\end{align*}
Hence the probability that the block is full ($B\geq t$) is at least $5/9$.
This is true for each of the $k$ blocks, so using linearity of expectation we have
$$
\frac{5k}{9} \leq \Exp[F] \leq \Pr[F\geq k/2]\cdot k + (1-\Pr[F\geq k/2])\cdot\frac{k}{2}.
$$
This implies $\Pr[F\geq k/2]\geq 1/9$, as claimed.
But then on all inputs with $|x|=8m$, $B$ outputs 1 with probability at least $\sigma/9$.

Algorithm $B$ uses $\alpha k\sqrt{tn}$ queries.
By~\cite{bbcmw:polynomialsj} and symmetrization, $B$'s acceptance probability is a
single-variate polynomial $p$ of degree $D\leq 2\alpha k\sqrt{tn}$ such that
\begin{quote}
$p(i)=0$ for all $i\in\{0,\ldots,m-1\}$,\\[1mm]
$p(8m)\geq\sigma/9$,\\[1mm]
$p(i)\in[0,1]$ for all $i\in\{0,\ldots,N\}$.
\end{quote}
The result now follows by applying \lemref{lem:keypol}
with $N=kn$, $m=kt/2$, $E=10$, and $\alpha$ a sufficiently small positive constant.
\end{proof}

\section{Time-Space Tradeoff for\\ Systems of Linear Inequalities}

Let $A$ be a fixed $N \times N$ matrix of nonnegative integers and let $x, b$ be
two input vectors of $N$ nonnegative integers smaller or equal to $t$.  A
\emph{matrix-vector product with upper bound}, denoted by $y = (A x)_{\le
b}$, is a vector $y$ such that $y_i = \min( (A x)[i], b_i)$.  An
\emph{evaluation of a system of linear inequalities $A x \ge b$}
is the $N$-bit vector of the truth values of the individual inequalities.
Here we present a quantum algorithm for matrix-vector product with
upper bound that satisfies time-space tradeoff $T^2 S = O(t N^3 (\log N)^5)$.
We then use our direct product theorems to show this is close to optimal.

\subsection{Upper bound}

It is easy to prove that matrix-vector products with upper bound $t$ can be
computed by a classical algorithm with $T S = O(N^2 \log t)$, as follows.  Let $S' = S / \log t$ and
divide the matrix $A$ into $(N/S')^2$ blocks of size $S' \times S'$ each.  The
output vector is evaluated row-wise as follows:  (1) Clear $S'$ counters,
one for each row, and read $b_i$.  (2) For each block, read $S'$ input
variables, multiply them by the corresponding submatrix of $A$, and update
the counters, but do not let them grow larger than $b_i$.  (3) Output the counters.  The space
used is $O(S' \log t) = O(S)$ and the total query complexity is $T = O(\frac N
{S'} \cdot \frac N {S'} \cdot S') = O(N^2 \log t / S)$.

The quantum algorithm \BMP{} works in a similar way and it is outlined in
\tabref{tab:alg-bmp}.  We compute the matrix product
in groups of $S' = S / \log N$ rows, read input variables, and update the
counters accordingly.  The advantage over the classical algorithm is that we
use the faster quantum search and quantum counting for finding non-zero entries.

The $u$-th row is called \emph{open} if its counter hasn't yet reached
$b_u$.  The subroutine \SMP{} maintains a set of open rows
$U \subseteq\{1,\ldots,S'\}$ and counters $0 \le y_u \le b_u$ for all $u \in U$.  We
process the input $x$ in blocks, each containing between $S' - O(\sqrt{S'})$
and $2 S' + O(\sqrt{S'})$
non-zero numbers at the positions $j$ where $A[u,j] \ne 0$ for some $u \in
U$.  The length $\ell$ of such a block is first found by iterated quantum counting 
(with number of queries specified in the proof below) and the non-zero input numbers 
are then found by a Grover search.  For each such number, we update 
all counters $y_u$ and close all rows that exceeded their threshold $b_u$.

\begin{table}[tb]
\noindent
\BMP{}
(fixed matrix $A_{N \times N}$, threshold $t$,
input vectors $x$ and $b$ of length $N$) \\
returns output vector $y = (A x)_{\le b}$:
\begin{itemize}
\item
For $i = 1, 2, \dots, \frac N {S'}$, where $S' = S / \log N$:
\begin{enumerate}
\item
Run \SMP{} on the $i$-th block of $S'$ rows of $A$.
\item
Output the $S'$ obtained results for those rows.
\end{enumerate}
\end{itemize}

\noindent
\SMP{}
(fixed $A_{S' \times N}$,
input $x_{N \times 1}$ and $b_{S' \times 1}$)
returns $y_{S' \times 1} = (A x)_{\le b}$:
\begin{enumerate}
\item
Initialize $y := (0, 0, \dots, 0)$, $p := 1$, $U := \{1,\ldots,S'\}$, and read
$b$.  Let $a_{1 \times N}$ denote an on-line computed row-vector with
$a_j = 1$ if $A[u, j]=1$ for some $u \in U$, and $a_j=0$ otherwise.
\item
While $p \le N$ and $U \ne \emptyset$, do the following:
\begin{enumerate}
\item
Let $\tilde{c}_{p,k}$ denote an estimate of the scalar product
\[
c_{p,k}=\sum_{j=p}^{p+k-1} a_j x_j.
\]
Initialize $k = S'$.
First, while $p + k-1 < N$ and $\tilde{c}_{p,k} < S'$, double $k$.
Second, find by binary search the maximal $\ell \in [\frac k 2, k]$ such that
$p + \ell-1 \le N$ and $\tilde{c}_{p,\ell} \le 2 S'$.
\item
Use quantum search to find the set $J$ of all positions $j \in [p, p+\ell-1]$
such that $a_j x_j > 0$.
\item
For all $j \in J$, read $x_j$,
and then do the following for all $u \in U$:
\begin{itemize}
\item
Increase $y_u$ by $A[u,j] x_j$.
\item
If $y_u \ge b_u$, set $y_u := b_u$ and remove $u$ from $U$.
\end{itemize}
\item
Increase $p$ by $\ell$.
\end{enumerate}
\item
Return $y$.
\end{enumerate}
\caption{Algorithm \BMP}
\label{tab:alg-bmp}
\end{table}

\begin{theorem}
\label{thm:matrixproduct}
\BMP{} has bounded error probability, its space complexity is $O(S)$,
and its query complexity is $T=O(N^{3/2} \sqrt t \cdot (\log N)^{5/2} / \sqrt {S})$.
\end{theorem}

\begin{proof}
The space complexity of \SMP{} is $O(S' \log N) = O(S)$,
because it stores a subset $U \subseteq\{1,\ldots,S'\}$, integer vectors $y, b$ of length
$S'$ with numbers at most $t \le N$, the set $J$ of size $O(S')$ with
numbers at most $N$, and a few counters.  Let us compute its query complexity.

Consider the $i$-th block found by \SMP{}; let $p_i$ be its
left column, let $\ell_i$ be its length, and let $U_i$ be the set of open rows
at the beginning of processing of this block.  The scalar product
$c_{p_i,\ell_i}$ is estimated by quantum counting
with $M = \sqrt{\ell_i}$ queries.  Finding a proper
$\ell_i$ requires $O(\log \ell_i)$ iterations.  Let $r_i$ be the number of
rows closed during processing of this block and let $s_i$ be the total
number added to the counters for other
(still open) rows in this block.  The numbers $\ell_i, r_i, s_i$ are random
variables.  If we instantiate them at the end of the quantum subroutine, the
following inequalities hold:
\[
\sum_i \ell_i	\le N, \quad
\sum_i r_i	\le S', \mbox{ and }\quad
\sum_i s_i	\le t S'.
\]
The iterated Grover search finds ones for two purposes: closing rows and
increasing counters.  Since each $b_i \le t$, the total cost in the $i$-th block is at most
\(
\sum_{j=1}^{r_i t} O(\sqrt{\ell_i / j})
  + \sum_{j=1}^{s_i} O(\sqrt{\ell_i / j})
  = O(\sqrt{\ell_i r_i t} + \sqrt{\ell_i s_i}).
\)
By a Cauchy-Schwarz inequality, the total number of queries that
\SMP{} spends in the Grover searches is at most
\begin{align*}
\sum_{i=1}^{\mathrm{\# blocks}} \hspace*{-0.5em}(\sqrt{\ell_i r_i t} + \sqrt{\ell_i s_i})
  & \le \sqrt{\sum_i \ell_i} \sqrt{t \sum_i r_i}
    + \sqrt{\sum_i \ell_i} \sqrt{\sum_i s_i}\\
 & \le \sqrt N \sqrt{t S'} + \sqrt N \sqrt{t S'}
  = O(\sqrt{N S' t}).
\end{align*}
The error probability of the Grover searches can be made polynomially small in
a logarithmic overhead.
It remains to analyze the outcome and error probability of quantum counting.
Let $c_i = c_{p_i,\ell_i} \in [S', 2 S']$.  One quantum
counting call with $M = \sqrt{\ell_i}$ queries gives an estimate $w$ such that
\[
|w - c_i| = O\left( \sqrt{c_i (\ell_i - c_i) \over \ell_i} + {\ell_i \over \ell_i} \right)
  = O(\sqrt c_i) = O(\sqrt{S'})
\]
with probability at least $8/\pi^2 \approx 0.8$.  We do it $O(\log N)$ times
and take the median, hence we obtain an estimate $\tilde{c}$ of $c_i$ with accuracy
$O(\sqrt{S'})$ with polynomially small error probability.  The result of
quantum counting is compared with the given threshold, that is with $S'$ or $2
S'$.  Binary search for
$\ell \in [\frac k 2, k]$ costs another factor of $\log k \le \log N$.  By a
Cauchy-Schwarz inequality, the total number of queries spent in the quantum
counting is at most $(\log N)^2$ times
\begin{align*}
\sum_i \sqrt{\ell_i}
  & \le \sqrt{\sum_i \ell_i} \sqrt{\sum_i 1} \le \sqrt N \sqrt{\#\rm blocks} \\
  & \le \sqrt N \sqrt{S' + t} \le \sqrt{N S' t},
\end{align*}
because in every block the algorithm closes a row or adds $\Theta(S')$ in total to
the counters.  The number of closed rows is at most $S'$ and the number $S'$
can be added at most $t$ times.

The total query complexity of \SMP{} is thus $O(\sqrt{N S'
t} \cdot (\log N)^2)$ and the query complexity of \BMP{} is
$N/S'$-times bigger.  The overall error probability is at most the sum of the
polynomially small error probabilities of the different subroutines, 
hence it can be kept below $1/3$.
\end{proof}

\subsection{Lower bound}

Here we use our direct product theorems to lower-bound the quantity $T^2S$ 
for $T$-query, $S$-space quantum algorithms for systems of linear inequalities.
The lower bound even holds if we fix $b$ to the all-$t$ vector $\vec{t}$
and let $A$ and $x$ be Boolean.

\begin{theorem}
Let $S=\min(O(N/t),o(N/\log N))$.
There exists an $N\times N$ Boolean matrix $A$ such that
every 2-sided error quantum algorithm that uses $T$ queries and $S$
qubits of space to decide a system $Ax\geq \vec{t}$ of $N$ inequalities, satisfies $T^2 S=\Omega(tN^3)$.
\end{theorem}

\begin{proof}
The proof is a modification of Theorem~22 of~\cite{ksw:dpt}
(quant-ph version). They use the probabilistic method to establish the following

\medskip

\noindent{\bf Fact:}
For every $k=o(N/\log N)$, there exists an $N\times N$ Boolean matrix $A$,
such that all rows of $A$ have weight $N/2k$, and every set of $k$ rows of $A$
contains a set $R$ of $k/2$ rows with the following property: each row in $R$ 
contains at least $n=N/6k$ ones that occur in no other row of $R$.

\medskip

Fix a matrix $A$ for $k=cS$, for some constant $c$ to be chosen later.
Consider a quantum circuit with $T$ queries and space $S$ that solves
the problem with success probability at least $2/3$.
We ``slice'' the quantum circuit into disjoint consecutive slices, each
containing $Q=\alpha \sqrt{tNS}$ queries, where $\alpha$ is the constant from
our direct product theorem (\thmref{thsdpt2sided}).
The total number of slices is $L=T/Q$.
Together, these disjoint slices contain all $N$ output gates.
Our aim below is to show that with sufficiently small constant $\alpha$
and sufficiently large constant $c$, no slice can produce more than $k$ outputs.
This will imply that the number of slices is $L\geq N/k$,
hence
$$
T=LQ\geq \frac{\alpha N^{3/2}\sqrt{t}}{c\sqrt{S}}.
$$

Now consider any slice. It starts with an $S$-qubit state
that is delivered by the previous slice and depends on the input,
then it makes $Q$ queries and outputs some $\ell$ results
that are jointly correct with probability at least $2/3$.
Suppose, by way of contradiction, that $\ell\geq k$.
Then there exists a set of $k$ rows of $A$ such that our slice produces
the $k$ corresponding results ($t$-threshold functions) with probability at least $2/3$.
By the above Fact, some set $R$ of $k/2$ of those rows has the property that each row in $R$
contains a set of $n=N/6k=\Theta(N/S)$ ones that do not occur in any of the $k/2-1$ other rows of $R$.
By setting all other $N-kn/2$ bits of $x$ to 0, we naturally get
that our slice, with the appropriate $S$-qubit starting state,
solves $k/2$ independent $t$-threshold functions $T_t$ on $n$ bits each.
(Note that we need $t\leq n/2=O(N/S)$; this follows from our assumption 
$S=O(N/t)$ with appropriately small constant in the $O(\cdot)$.)
Now we replace the initial $S$-qubit state by the completely
mixed state, which has ``overlap'' $2^{-S}$ with every $S$-qubit state. 
This turns the slice into a stand-alone algorithm
solving $\vectorres{T_t}{k/2}$ with success probability
$$
\sigma\geq\frac{2}{3}2^{-S}.
$$
But this algorithm uses only $Q=\alpha \sqrt{tNS}=O(\alpha k \sqrt{tn})$
queries, so our direct product theorem 
(\thmref{thsdpt2sided})
with sufficiently small constant $\alpha$ implies
$$
\sigma\leq 2^{-\Omega(k/2)}=2^{-\Omega(cS/2)}.
$$
Choosing $c$ a sufficiently large constant (independent of this
specific slice), our upper and lower bounds on $\sigma$ contradict.
Hence the slice must produce fewer than $k$ outputs.
\end{proof}

It is easy to see that the case $S\geq N/t$ (equivalently, $t\geq N/S$)
is at least as hard as the  $S=N/t$ case, for which we have the lower bound 
$T^2S=\Omega(t N^3)=\Omega(N^4/S)$, hence $TS=\Omega(N^2)$. 
But that lower bound matches the \emph{classical}
deterministic upper bound up to a logarithmic factor
and hence is essentially tight also for quantum.
We thus have two different regimes for space: for small space, a quantum
computer is faster than a classical one in solving systems of linear
inequalities, while for large space it is not.

A similar slicing proof using \thmref{thsdpt1sided}
(with each slice of $Q=\alpha \sqrt{NS}$ queries producing at most $S/t$ outputs) 
gives the following lower bound on time-space tradeoffs for 1-sided error algorithms.

\begin{theorem}
Let $t\leq S\leq \min(O(N/t^2),o(N/\log N))$.
There exists an $N\times N$ Boolean matrix $A$ such that every 1-sided
error quantum algorithm that uses $T$ queries and $S$ qubits of space
to decide a system $Ax\geq \vec{t}$ of $N$ inequalities, satisfies $T^2S=\Omega(t^2 N^3)$.
\end{theorem}

Note that our lower bound $\Omega(t^2 N^3)$ for 1-sided error algorithms is higher 
by a factor of $t$ than the best upper bounds for 2-sided error algorithms.
This lower bound is probably not optimal.
If $S>N/t^2$ then the essentially optimal classical tradeoff $TS=\Omega(N^2)$ takes over.

\section{Summary}

In this paper we described a new version of the adversary method for quantum query lower bounds,
based on analyzing the eigenspace structure of the problem we want to lower bound.
We proved two new quantum direct product theorems, the first using the new adversary method, 
the second using the polynomial method:
\begin{itemize}
\item For every symmetric function $f$,
every 2-sided error quantum algorithm for $\vectorres{f}{k}$
using fewer than $\alpha k Q_2(f)$ queries has success probability at most $2^{-\Omega(k)}$.
\item For every $t$-threshold function $f$, every 1-sided error quantum algorithm for $\vectorres{f}{k}$
using fewer than $\alpha k Q_2(f)$ queries has success probability at most $2^{-\Omega(kt)}$.
\end{itemize}
Both results are tight up to constant factors.
{}From these results we derived the following time-space tradeoffs for quantum algorithms that
decide a system $Ax\geq b$ of $N$ linear inequalities (where $A$ is a fixed
$N\times N$ matrix of nonnegative integers, $x,b$ are variable, and $b_i\leq t$ for all $i$):
\begin{itemize}
\item Every $T$-query, $S$-space 2-sided error quantum algorithm for evaluating $Ax\geq b$
satisfies $T^2S=\Omega(tN^3)$ if $S\leq N/t$, and satisfies $TS=\Omega(N^2)$ if $S>N/t$.
We gave an algorithm matching these bounds up to polylog factors.
\item Every $T$-query, $S$-space 1-sided error quantum algorithm for evaluating $Ax\geq b$
satisfies $T^2S=\Omega(t^2N^3)$ if $t\leq S\leq N/t^2$, and satisfies $TS=\Omega(N^2)$ if $S>N/t^2$.
We do not have a matching algorithm in the first case and conjecture that this bound is not tight.
\end{itemize}

\bibliographystyle{abbrv}

\appendix

\section{Proofs from Section 3}		
\label{app:andrisproof}

\begin{table*}[tb]
\[
\begin{tabular}{|l l|}
\hline
$\ket{\psi^a_{i_1,\dots,i_j}}$ &
  uniform superposition of states with $|x| = t-1+a$ and with $j$ fixed bits
  set to 1 \\
$T_{j,a}$ &
  spanned by $\ket{\psi^a_{i_1,\dots,i_j}}$ for all $j$-tuples $i$ \\
$S_{j,a} = T_{j,a} \cap T_{j-1,a}^\perp$ &
  that is, we remove the lower-dimensional subspace \\
$\ket{\tilde\psi^a_{i_1,\dots,i_j}}$ &
  projection of $\ket{\psi^a_{i_1,\dots,i_j}}$ onto $S_{j,a}$ \\
$S_{j,\pm}$ &
  spanned by $\frac{\ket{\tilde\psi^0}}{\|\tilde\psi^0\|} \pm
  \frac{\ket{\tilde\psi^1}}{\|\tilde\psi^1\|}$ \\
$R_j = S_{j,+}$ &
  for $j < \frac t 2$
  \hfill \dots bad subspaces \kern 100pt \\
$R_{t/2}$ &
  direct sum of $S_{j,+}$ for $j \ge t/2$, and all $S_{j,-}$
  \hfill \dots good subspaces \kern 100pt \\
\hline
$\S_{m-} = \bigoplus\limits_{\substack{|r|=m\\j}} \bigotimes\limits_{i=1}^k S_{j_i,r_i}$ &
  where $|r|$ is the number of minuses in $r=r_1,\ldots,r_k$ \\
$\R_m = \bigoplus\limits_{|j|_1=m} \bigotimes\limits_{i=1}^k R_{j_i}$ &
  where $|j|_1$ is the sum of all entries in $j=j_1,\ldots,j_k$\\
$\R'_j = \bigoplus\limits_{m \ge j} \R_m$ &
  \\
\hline
$\ket{\psi^{a,b}_{i_1,\dots,i_j}}$ &
  uniform superposition of states with $|x| = t-1+a$, with $j$ fixed bits set
  to 1, and $x_1 = b$ \\
$T_{j,a,b}$ &
  spanned by $\ket{\psi^{a,b}_{i_1,\dots,i_j}}$ for all $j$-tuples $i$ \\
$S_{j,a,b} = T_{j,a,b} \cap T_{j-1,a,b}^\perp$ &
  that is, we remove the lower-dimensional subspace \\
$\ket{\tilde\psi^{a,b}_{i_1,\dots,i_j}}$ &
  projection of $\ket{\psi^{a,b}_{i_1,\dots,i_j}}$ into $S_{j,a,b}$ \\
$S_{j,a}^{\alpha,\beta}$ &
  spanned by $\alpha \frac{\ket{\tilde\psi^{a,0}}}{\|\tilde\psi^{a,0}\|} +
  \beta \frac{\ket{\tilde\psi^{a,1}}}{\|\tilde\psi^{a,1}\|}$ \\
\hline
\end{tabular}
\]
\caption{States and subspaces used in the proof}
\label{tab:states}
\end{table*}

\subsection{Proof of Lemma \ref*{lem:prob}}
\label{app:proof1}

The measurement of $\H_A$ decomposes the state in the $\H_I$ register as
follows: 
\[ \rho=\sum_{a_1, \ldots, a_k\in\{0, 1\}} p_{a_1, \ldots, a_k} 
\sigma_{a_1, \ldots, a_k} ,\]
with $p_{a_1, \ldots, a_k}$ being the probability of the measurement giving the
answer $(a_1, \ldots, a_k)$ (where $a_j=1$ means the algorithm outputs---not 
necessarily correctly---that $|x^j|=t$ and $a_j=0$ means $|x^j|=t-1$)
and $\sigma_{a_1, \ldots, a_k}$ being the density matrix
of $\H_I$, conditional on this outcome of the measurement.
Since the support of $\rho$ is contained in 
$\S_{0-}\oplus \cdots \oplus \S_{m-}$,
the support of the states $\sigma_{a_1, \ldots, a_k}$ is also contained in 
$\S_{0-}\oplus \cdots \oplus \S_{m-}$.
The probability that the answer $(a_1, \ldots, a_k)$ is correct
is equal to
\begin{equation}
\label{eq:tobound} 
\Tr \P_{\otimes_{j=1}^k \oplus_{l=0}^{t-1+a_j} S_{l, a_j}} \sigma_{a_1, \ldots, a_k} .
\end{equation}
We show that, for any $\sigma_{a_1, \ldots, a_k}$ with support
contained in $\S_{0-}\oplus \cdots\oplus \S_{m-}$, \eqnref{eq:tobound}
is at most $\frac{\sum_{m'=0}^m {k \choose m'}}{2^k}$.

For brevity, we now  
write $\sigma$ instead of $\sigma_{a_1, \ldots, a_k}$.
A measurement w.r.t.\ $\otimes_{j=1}^k \oplus_{l} S_{l, a_j}$ and 
its orthogonal complement commutes with a measurement
w.r.t. the collection of subspaces
\[ \otimes_{j=1}^k (S_{l_j, 0}\oplus S_{l_j, 1}) ,\]
where $l_1, \ldots, l_k$ range over $\{0, \ldots, t\}$.
Therefore 
\[ \Tr \P_{\otimes_{j=1}^k \oplus_{l} S_{l, a_j}} \sigma =
 \sum_{l_1, \ldots, l_k} \Tr \P_{\otimes_{j=1}^k \oplus_{l} S_{l, a_j}}
\P_{\otimes_{j=1}^k (S_{l_j, 0}\oplus S_{l_j, 1})} \sigma .\]
Hence to bound \eqnref{eq:tobound} it 
suffices to prove the same bound with 
\[ \sigma' = \P_{\otimes_{j=1}^k (S_{l_j, 0}\oplus S_{l_j, 1})} \sigma .\]
instead of $\sigma$.
Since 
\[ \left( \otimes_{j=1}^k (S_{l_j, 0}\oplus S_{l_j, 1}) \right) \cap 
\left( \otimes_{j=1}^k (\oplus_{l} S_{l, a_j}) \right) = 
\otimes_{j=1}^k S_{l_j, a_j} ,\]
we have 
\begin{equation}
\label{eq:tobound1} 
\Tr \P_{\otimes_{j=1}^k (\oplus_{l} S_{l, a_j})} \sigma' = 
\Tr \P_{\otimes_{j=1}^k S_{l_j, a_j}} \sigma' .
\end{equation}
We prove this bound for the case when $\sigma'$ is a pure state:
$\sigma'=\ket{\psi}\bra{\psi}$. Then equation~\eqnref{eq:tobound1}
is equal to 
\begin{equation}
\label{eq:tobound2}  
\| \P_{\otimes_{j=1}^k S_{l_j, a_j}} \psi \|^2 .
\end{equation}
The bound for mixed states $\sigma'$ follows by decomposing
$\sigma'$ as a mixture of pure states $\ket{\psi}$, bounding
\eqnref{eq:tobound2} for each of those states and then summing up the bounds.

We have
\[ (\S_{0-}\oplus \cdots \oplus \S_{m-}) \cap 
(\bigotimes_{j=1}^k  (S_{l_j, 0}\oplus S_{l_j, 1})) =
\hspace{-1em}\mathop{\bigoplus_{r_1, \ldots, r_k\in\{+, -\},}}_{|\{ i:r_i=-\}|\leq m} 
\bigotimes_{j=1}^k S_{l_j, r_j} .\]
We express
\[ \ket{\psi} = \mathop{\sum_{r_1, \ldots, r_k\in\{+, -\},}}_{
|\{ i:r_i=-\}|\leq m}
\alpha_{r_1, \ldots, r_k} \ket{\psi_{r_1, \ldots, r_k}} ,\]
with $\ket{\psi_{r_1, \ldots, r_k}} \in \otimes_{j=1}^k S_{l_j, r_j}$.
Therefore 
\begin{align}
\hspace{-.5em}\| \P_{\otimes_{j=1}^k S_{l_j, a_j}} \psi \|^2 &\leq
\left(\hspace{-.2em} \sum_{r_1, \ldots, r_k} \hspace{-.5em}|\alpha_{r_1, \ldots, r_k}| 
\|\P_{\otimes_{j=1}^k S_{l_j, a_j}} \psi_{r_1, \ldots, r_k}\| 
\right)^2 \nonumber \\
&\le \label{eq:pmk} 
\sum_{r_1, \ldots, r_k}
\|\P_{\otimes_{j=1}^k S_{l_j, a_j}} \psi_{r_1, \ldots, r_k}\|^2 , 
\end{align}
where the second inequality uses Cauchy-Schwarz and
\[ \|\psi\|^2=\sum_{r_1, \ldots, r_k} |\alpha_{r_1, \ldots, r_k}|^2 =1. \]

\begin{claim}
\label{claim:plusminus}
$\displaystyle \|\P_{\otimes_{j=1}^k S_{l_j, a_j}} \psi_{r_1, \ldots, r_k}\|^2 \leq \frac{1}{2^k}.$
\end{claim}

\begin{proof}
Let $\ket{\varphi^{j, 0}_i}$, $i\in[\dim S_{l_j, 0}]$ form a basis for 
the subspace $S_{l_j, 0}$. Define a map 
$U_j:S_{l_j, 0}\rightarrow S_{l_j, 1}$
by $U_j\ket{\tilde{\psi}^{0}_{i_1, \ldots, i_{l_j}}}=
\ket{\tilde{\psi}^{1}_{i_1, \ldots, i_{l_j}}}$. 
Then $U_j$ is a multiple of a
unitary transformation: $U_{j}=c_{j} U'_{j}$ for some
unitary $U'_{j}$ and a constant $c_{j}$.
(This follows from \clmref{claim:unitary} in \appref{app:proof2}.)

Let $\ket{\varphi^{j, 1}_i}=U'_j\ket{\varphi^{j, 0}_i}$.
Since $U'_j$ is a unitary
transformation, the states $\ket{\varphi^{j, 1}_i}$ form a basis 
for $S_{l_j, 1}$.
Therefore 
\begin{equation}
\label{eq:basis} 
\bigotimes_{j=1}^k \ket{\varphi^{j, a_j}_{i_j}} 
\end{equation}
is a basis for $\otimes_{j=1}^k S_{l_j, a_j}$.
Moreover, the states
\[ \ket{\varphi^{j, +}_i}=\frac{1}{\sqrt{2}} \ket{\varphi^{j, 0}_i} +  
\frac{1}{\sqrt{2}} \ket{\varphi^{j, 1}_i}, \mbox{~~}
\ket{\varphi^{j, -}_i}=\frac{1}{\sqrt{2}} \ket{\varphi^{j, 0}_i} -  
\frac{1}{\sqrt{2}} \ket{\varphi^{j, 1}_i} \]
are a basis for $S_{l_j, +}$ and $S_{l_j, -}$, respectively.
Therefore
\begin{equation}
\label{eq:sum2n}
 \ket{\psi_{r_1, \ldots, r_k}} = \sum_{i_1, \ldots, i_k}
\alpha_{i_1, \ldots, i_k} \bigotimes_{j=1}^k 
\ket{\varphi^{j, r_j}_{i_j}} .
\end{equation}
The inner product between $\otimes_{i=1}^k \ket{\varphi^{j, a_j}_{i'_j}}$
and $\otimes_{j=1}^k \ket{\varphi^{j, r_j}_{i_j}}$ is 
\[ \prod_{j=1}^k \bra{\varphi^{j, r_j}_{i_j}} \varphi^{j, a_j}_{i'_j}\rket .\]
Note that $r_j \in \{ +, - \}$ and $a_j \in \{ 0, 1 \}$.
The terms in this product are $\pm \frac{1}{\sqrt{2}}$ if 
$i'_j=i_j$ and 0 otherwise. 
This means that $\otimes_{j=1}^k \ket{\varphi^{j, r_j}_{i_j}}$ has 
inner product $\pm \frac{1}{2^{k/2}}$ with
$\otimes_{i=1}^k \ket{\varphi^{j, a_j}_{i_j}}$ and inner
product 0 with all other basis states \eqnref{eq:basis}.
Therefore, 
\[ \P_{\otimes_{j=1}^k S_{l_j, a_j}} \otimes_{j=1}^k \ket{\varphi^{j, r_j}_{i_j}} =
\pm \frac{1}{2^{k/2}}  \otimes_{i=1}^k \ket{\varphi^{j, a_j}_{i_j}} .\]
Together with equation \eqnref{eq:sum2n}, this means that
\[ \|\P_{\otimes_{j=1}^k S_{l_j, a_j}} \psi_{r_1, \ldots, r_k}\| \leq \frac{1}{2^{k/2}}
\| \psi_{r_1, \ldots, r_k}\| = \frac{1}{2^{k/2}} .\]
Squaring both sides completes the proof of the claim.
\end{proof}

Since there are ${k \choose m'}$ tuples $(r_1, \ldots, r_k)$
with $r_1, \ldots, r_k \in\{+, -\}$ and 
$|\{i:r_i=-\}|=m'$, 
\clmref{claim:plusminus} together with equation \eqnref{eq:pmk}
implies
\[ 
\| \P_{\otimes_{j=1}^k S_{l_j, a_j}} \psi \|^2 \leq
\frac{\sum_{m'=0}^m {k \choose m'}}{2^k}.
\] 

\subsection{Proof of Corollary \ref*{cor:prob}}
\label{app:proof1cor}

Let $\ket{\psi}$ be a purification of $\rho$ in $\H_A\otimes\H_I$.
Let
\[ \ket{\psi}= \sqrt{1-\delta}\ket{\psi'}+\sqrt{\delta}\ket{\psi''} \]
where $\ket{\psi'}$ is in the subspace 
$\H_A \otimes(\S_{0-}\oplus \S_{1-}\oplus \cdots \oplus \S_{m-})$ and 
$\ket{\psi''}$ is in the subspace 
$\H_A \otimes (\S_{0-}\oplus \S_{1-}\oplus \cdots\oplus \S_{m-})^{\perp}$.
Then $\delta=\Tr \P_{(\S_{0-}\oplus \cdots \oplus \S_{m-})^{\perp}} \rho$.

The success probability of $\A$ is the probability that, if we measure both
the register $\H_A$ containing the result of the computation and $\H_I$,
then we get $a_1, \ldots, a_k$ and $x^1, \ldots, x^k$ such that
$x^j$ contains $t-1+a_j$ ones for every $j\in\{1, \ldots, k\}$. 

Consider the probability of getting $a_1,\ldots,a_k\in\01$ and 
$x^1,\ldots,x^k\in\01^n$ with this property, when measuring $\ket{\psi'}$ 
(instead of $\ket{\psi}$). By \lemref{lem:prob}, this probability is
at most $\frac{\sum_{m'=0}^m {k \choose m'}}{2^k}$.
We have 
\begin{align*} 
\|\psi -\psi'\| & \leq (1-\sqrt{1-\delta}) 
\|\psi'\| + \sqrt{\delta} \|\psi''\| \\
 & = (1-\sqrt{1-\delta}) + \sqrt{\delta} \leq 2 \sqrt{\delta} .
\end{align*}
We now apply

\begin{lemma}
[\cite{bernstein&vazirani:qcomplexity}]
\label{lem:bv}
For any states $\ket{\psi}$ and $\ket{\psi'}$ and any measurement $M$,
the variational distance between the 
probability distributions obtained by applying $M$ to $\ket{\psi}$
and $\ket{\psi'}$ is at most $2\|\psi-\psi'\|$.
\end{lemma}

Hence the success probability of $\A$ is at most
\[ \frac{\sum_{m'=0}^m {k \choose m'}}{2^k}+4\sqrt{\delta} =
\frac{\sum_{m'=0}^m {k \choose m'}}{2^k}+
4\sqrt{\Tr \P_{(\S_{0-}\oplus \cdots \oplus \S_{m-})^{\perp}} \rho} .\]

\subsection{Structure of the subspaces when asking one query}

Let $\ket{\psi_d}$ be the state of $\H_A\otimes \H_I$ after $d$ queries.
Write
\[ \ket{\psi_d} =\sum_{i=0}^{k n} a_i \ket{\psi_{d, i}},\]
with $\ket{\psi_{d, i}}$ being the part 
in which the query register contains $\ket{i}$.
Let $\rho_{d, i}=\Tr_{\H_A}\ket{\psi_{d, i}}\bra{\psi_{d, i}}$. 
Then 
\begin{equation}
\label{eq-2507a}
\rho_d=\sum_{i=0}^{k n} a^2_i \rho_{d, i}.
\end{equation}
Because of
\[ \Tr \P_{\R_{m}}\rho_d = \sum_{i=0}^{k n} a^2_i 
\Tr \P_{\R_{m}}\rho_{d, i} ,\]
we have $P(\rho_{d}) = \sum_{i=0}^{k n} a^2_i P(\rho_{d, i})$.
Let $\rho'_d$ be the state after the $d$-th query and
let $\rho'_d=\sum_{i=0}^{k n} a^2_i \rho'_{d, i}$ be
a decomposition similar to equation \eqnref{eq-2507a}.
\lemref{lem:onestep} follows by showing
\begin{equation}
\label{eq:change1} 
P(\rho'_{d, i}) \leq  \left(1+ \frac{C}{\sqrt{t n}} (q^{t/2}-1) + 
\frac{C\sqrt{t}}{\sqrt{n}} (q-1) \right) P(\rho_{d, i})  
\end{equation}
for each $i$.
For $i=0$, the query does not change the state if the query register
contains $\ket{i}$. Therefore, 
$\rho'_{d, 0}=\rho_{d, 0}$ and $P(\rho'_{d, 0})=P(\rho_{d, 0})$.
This means that equation \eqnref{eq:change1} is true for $i=0$.
To prove the $i\in\{1, \ldots, k n\}$ case, it suffices to prove 
the $i=1$ case (because of symmetry).

Let $\ket{\psi^{a, b}_{i_1, \ldots, i_j}}$ (with $a, b\in\{0, 1\}$
and $i_1, \ldots, i_j \in \{2, \ldots, n\}$)
be the uniform superposition over basis states
$\ket{b, x_2, \ldots, x_n}$ 
(of $\H_{one}$) with $b+x_2+\cdots+x_n=t-1+a$
and $x_{i_1}=\cdots=x_{i_j}=1$.
Let $T_{j, a, b}$  be the space spanned by all states
$\ket{\psi^{a, b}_{i_1, \ldots, i_j}}$ 
and let $S_{j, a, b}=T_{j, a, b}\cap T_{j-1, a, b}^\perp$.
Let $\ket{\tilde{\psi}^{a, b}_{i_1, \ldots, i_j}} = \P_{T_{j-1, a, b}^\perp}
\ket{\psi^{a, b}_{i_1, \ldots, i_j}}$.

Let $S_{j,a}^{\alpha,\beta}$ be the subspace spanned by all states
\begin{equation}
\alpha\frac{\ket{\tilde{\psi}^{a,0}_{i_1, \ldots, i_j}}}{
\|\tilde{\psi}^{a,0}_{i_1, \ldots, i_j}\|}+
\beta \frac{\ket{\tilde{\psi}^{a,1}_{i_1, \ldots, i_j}}}{
\|\tilde{\psi}^{a,1}_{i_1, \ldots, i_j}\|} .
\label{eq-2507}
\end{equation}

\begin{claim}
\label{clm:relate}
Let $\alpha_a=\sqrt{\frac{n-(t-1+a)}{n-j}} 
\|\tilde{\psi}^{a, 0}_{i_1, \ldots, i_j}\|$ and
$\beta_a=\sqrt{\frac{(t-1+a)-j}{n-j}} \|\tilde{\psi}^{a, 1}_{i_1, \ldots, i_j}\|$.
Then
(i)  $S_{j,a}^{\alpha_a, \beta_a}\subseteq S_{j,a}$
and
(ii)  $S_{j,a}^{\beta_a, -\alpha_a}\subseteq S_{j+1,a}$.
\end{claim}

\begin{proof}
For part (i), consider the states $\ket{\psi^a_{i_1, \ldots, i_j}}$ 
in $T_{j,a}$, for $1\not\in\{i_1,\ldots,i_j\}$.  We have
\begin{align}
\ket{\psi^a_{i_1, \ldots, i_j}} 
&=\sqrt{\frac{n-(t-1+a)}{n-j}} \ket{\psi^{a,0}_{i_1, \ldots, i_j}} \nonumber \\
&+\sqrt{\frac{(t-1+a)-j}{n-j}} \ket{\psi^{a,1}_{i_1, \ldots, i_j}} 
\label{eq-new} 
\end{align}
because among the states $\ket{x_1\ldots x_n}$ with
$|x|=t-1+a$ and $x_{i_1}=\cdots=x_{i_j}=1$, 
a $\frac{n-(t-1+a)}{n-j}$ fraction have $x_1=0$ and the rest have $x_1=1$.
The projections of these states to $T_{j-1,a,0}^\perp \cap T_{j-1,a,1}^\perp$
are
\[
\sqrt{\frac{n-(t-1+a)}{n-j}} \ket{\tilde{\psi}^{a,0}_{i_1, \ldots, i_j}}+
\sqrt{\frac{(t-1+a)-j}{n-j}} \ket{\tilde{\psi}^{a,1}_{i_1, \ldots, i_j}} 
\]
which, by equation (\ref{eq-2507}) are exactly the states spanning
$S^{\alpha_a, \beta_a}_{j,a}$.
Furthermore, we claim that 
\begin{equation}
\label{eq-2707} T_{j-1,a}\subseteq T_{j-1,a,0}\oplus T_{j-1,a,1} \subseteq
T_{j,a}. 
\end{equation}
The first containment is true because $T_{j-1,a}$ is spanned by 
the states $\ket{\psi^a_{i_1, \ldots, i_{j-1}}}$
which either belong to $T_{j-2,a,1}\subseteq T_{j-1,a,1}$
(if $1\in\{i_1, \ldots, i_{j-1}\}$)
or are a linear combination of states $\ket{\psi^{a,0}_{i_1, \ldots, i_{j-1}}}$
and $\ket{\psi^{a,1}_{i_1, \ldots, i_{j-1}}}$ (by equation~(\ref{eq-new})), 
which belong to $T_{j-1,a,0}$ and $T_{j-1,a,1}$.
The second containment follows because the
states $\ket{\psi^{a, 1}_{i_1, \ldots, i_{j-1}}}$ spanning
$T_{j-1,a,1}$ are the same as the states 
$\ket{\psi^a_{1, i_1, \ldots, i_{j-1}}}$ which belong to $T_{j,a}$, 
and the states $\ket{\psi^{a, 0}_{i_1, \ldots, i_{j-1}}}$ spanning
$T_{j-1,a,0}$ can be expressed as linear combinations
of $\ket{\psi^a_{i_1, \ldots, i_{j-1}}}$ and
$\ket{\psi^a_{1, i_1, \ldots, i_{j-1}}}$ which both belong to $T_{j,a}$.

The first part of (\ref{eq-2707}) now implies 
\[ S^{\alpha_a, \beta_a}_{j,a}\subseteq T_{j-1,a,0}^\perp \cap T_{j-1,a,1}^\perp \subseteq
T_{j-1,a}^{\perp} .\]
Also,
$S^{\alpha_a, \beta_a}_{j,a}\subseteq T_{j,a}$,
because $S^{\alpha_a, \beta_a}_{j,a}$
is spanned by the states
\begin{align*}
\P_{T_{j-1,a,0}^\perp \cap T_{j-1,a,1}^\perp}& \ket{\psi^a_{i_1, \ldots, i_j}} \\
& = \ket{\psi^a_{i_1, \ldots, i_j}} - \P_{T_{j-1,a,0} \oplus T_{j-1,a,1}} 
\ket{\psi^a_{i_1, \ldots, i_j}} 
\end{align*}
and $\ket{\psi^a_{i_1, \ldots, i_j}}$ belongs to $T_{j,a}$ by the definition
of $T_{j,a}$ and 
$\P_{T_{j-1,a,0} \oplus T_{j-1,a,1}} \ket{\psi^a_{i_1, \ldots, i_j}}$ belongs to
$T_{j,a}$ because of the second part of (\ref{eq-2707}).
Therefore, 
$S^{\alpha_a, \beta_a}_{j,a} \subseteq T_{j,a} \cap T_{j-1,a}^{\perp}=S_{j,a}$.

For part (ii), 
we have
\[ S^{\alpha_a, \beta_a}_{j,a}\subseteq S_{j,a,0}\oplus S_{j,a,1}
\subseteq T_{j,a,0} \oplus T_{j,a,1} \subseteq T_{j+1,a} ,\]
where the first containment is true because $S^{\alpha_a, \beta_a}_{j,a}$
is spanned by linear combinations of vectors
$\ket{\tilde{\psi}^{a,0}_{i_1, \ldots, i_j}}$ (which belong to $S_{j,a,0}$)
and vectors $\ket{\tilde{\psi}^{a,1}_{i_1, \ldots, i_j}}$ 
(which belong to $S_{j,a,1}$) and the last containment is
true because of the second part of equation (\ref{eq-2707}).
Now let
\begin{equation}
\label{eq-2707a} 
\ket{\psi}=\beta_a \frac{\ket{\tilde{\psi}^{a,0}_{i_1, \ldots, i_j}}}{
\|\tilde{\psi}^{a,0}_{i_1, \ldots, i_j}\| }
-\alpha_a \frac{\ket{\tilde{\psi}^{a,1}_{i_1, \ldots, i_j}}}{
\|\tilde{\psi}^{a,1}_{i_1, \ldots, i_j}\|} 
\end{equation}
be one of the vectors spanning $S^{\beta_a, -\alpha_a}_{j,a}$.
To prove that $\ket{\psi}$ is in $S_{j+1,a}=T_{j+1,a} \cap T_{j,a}^\perp$, it remains to
prove that $\ket{\psi}$ is orthogonal to $T_{j,a}$.
This is equivalent to proving that $\ket{\psi}$ is orthogonal to
each of the vectors $\ket{\psi^a_{i'_1, \ldots, i'_j}}$ spanning $T_{j,a}$.
We distinguish two cases (note that $1\not\in\{i_1, \ldots, i_j\}$):

\noindent
{\bf Case 1.} $1\in\{i'_1, \ldots, i'_j\}$.

For simplicity, assume $1=i'_j$. Then
$\ket{\psi^a_{i'_1, \ldots, i'_j}}$ is the same 
as $\ket{\psi^{a, 1}_{i'_1, \ldots, i'_{j-1}}}$,
which belongs to $T_{j-1,a,1}$. By definition,
the vector $\ket{\psi}$ belongs to $T_{j-1,a,0}^\perp \cap T_{j-1,a,1}^\perp$
and is therefore orthogonal to $\ket{\psi^{a, 1}_{i'_1, \ldots, i'_{j-1}}}$. 

\noindent
{\bf Case 2.} $1\not\in\{i'_1, \ldots, i'_j\}$.

We will prove this case by induction on $\ell=|\{i'_1,\ldots,i'_j\}-\{i_1,\ldots,i_j\}|$.

In the base step $(\ell=0$), we have $\{i'_1,\ldots,i'_j\}=\{i_1,\ldots,i_j\}$.
Since $\ket{\psi}$ belongs to $T_{j-1,a,0}^\perp \cap T_{j-1,a,1}^\perp$,
it suffices to prove $\ket{\psi}$ is orthogonal to the projection
of  $\ket{\psi^a_{i_1, \ldots, i_j}}$ to $T_{j-1,a,0}^\perp \cap T_{j-1,a,1}^\perp$
which, by the discussion after equation (\ref{eq-new}), 
equals 
\begin{equation}
\label{eq-2707b} 
\alpha_a \frac{\ket{\tilde{\psi}^{a,0}_{i_1, \ldots, i_j}}}{
\|\tilde{\psi}^{a,0}_{i_1, \ldots, i_j}\| }
+\beta_a \frac{\ket{\tilde{\psi}^{a,1}_{i_1, \ldots, i_j}}}{
\|\tilde{\psi}^{a,1}_{i_1, \ldots, i_j}\|} .
\end{equation}
{}From equations (\ref{eq-2707a}) and (\ref{eq-2707b}), we see that the
inner product of the two states is $\alpha_a \beta_a -\beta_a \alpha_a=0$.

For the inductive step ($\ell\geq 1$), assume $i'_j\not\in\{i_1, \ldots, i_j \}$.
Up to renormalization, we have
\[ \ket{\psi^a_{i'_1, \ldots, i'_{j-1}}}= 
\sum_{i'\notin\{i'_1, \ldots, i'_{j-1}\}} 
\ket{\psi^a_{i'_1, \ldots, i'_{j-1}, i'}} .\]
Because $\ket{\psi^a_{i'_1, \ldots, i'_{j-1}}}$ is in 
$T_{j-1,a,0}\oplus T_{j-1,a,1}$, we have
\begin{equation}\label{eq-new0}
\sum_{i'\notin\{i'_1, \ldots, i'_{j-1}\}} 
\bra{\psi^a_{i'_1, \ldots, i'_{j-1}, i'}} \psi\rket=
\bra{\psi^a_{i'_1, \ldots, i'_{j-1}}} \psi\rket=0.
\end{equation}
As proven in the previous case, $\lbra \psi^a_{i'_1, \ldots, i'_{j-1},1} \ket{\psi}=0$.
Moreover, by the induction hypothesis we have $\lbra \psi^a_{i'_1, \ldots, i'_{j-1},i'} \ket{\psi}=0$
whenever $i'\in\{i_1, \ldots, i_j\}$.
Therefore equation (\ref{eq-new0}) reduces to
\begin{equation}
\label{eq-new1} 
\sum_{i'\notin\{i'_1, \ldots, i'_{j-1},i_1,\ldots,i_j, 1\}} 
\bra{\psi^a_{i'_1, \ldots, i'_{j-1}, i'}} \psi\rket
=0 .
\end{equation}
By symmetry, the inner products in this sum are
the same for every $i'$. Hence they are all 0, in particular for $i'=i'_j$.
\end{proof}

\subsection{Proof of Lemma \ref*{lem:onestep}}
\label{app:proof2}

\begin{claim}
\label{claim:unitary}
The maps $U_{01}:S_{j, 0, 0}\rightarrow S_{j, 0, 1}$,
$U_{10}:S_{j, 0, 0}\rightarrow S_{j, 1, 0}$ and 
$U_{11}:S_{j, 0, 0}\rightarrow S_{j, 1, 1}$ defined by
$U_{ab}\ket{\tilde{\psi}^{0, 0}_{i_1, \ldots, i_j}}=
\ket{\tilde{\psi}^{a, b}_{i_1, \ldots, i_j}}$ are multiples of
unitary transformations: $U_{ab}=c_{ab} U'_{ab}$ for some
unitary $U'_{ab}$ and some constant $c_{ab}$.
\end{claim}

\begin{proof}
We define $M:T_{j, 0, 0}\rightarrow T_{j, 0, 1}$ by 
\[ M\ket{0 x_2 \ldots x_n}=\sum_{\ell:x_{\ell}=1} 
\ket{1 x_2 \ldots x_{\ell-1} 0 x_{\ell+1} \ldots x_n} .\]
Note that $M$ does not depend on $j$.
We claim
\begin{align}
\label{eq:map}
M\ket{\tilde{\psi}^{0, 0}_{i_1, \ldots, i_j}} & = c \ket{\tilde{\psi}^{0, 1}_{i_1, \ldots, i_j}},\\
M^{\dagger} \ket{\tilde{\psi}^{0, 1}_{i_1, \ldots, i_j}} & = c' \ket{\tilde{\psi}^{0, 0}_{i_1, \ldots, i_j}},\nonumber
\end{align}
for some constants $c$ and $c'$ that may depend on $n, t$ and $j$ but not on $i_1, \ldots, i_j$.
To prove that, we need to prove two things.  First, we claim that
\begin{equation}
\label{eq-rem4}
M\ket{\psi^{0,0}_{i_1, \ldots, i_j}}=c \ket{\psi^{0,1}_{i_1, \ldots, i_j}}+\ket{\psi'},
\end{equation}
where $\ket{\psi'}\in T_{j-1,0,1}$ (note that $1\not\in\{i_1,\ldots,i_j\}$).
Equation~\eqnref{eq-rem4} follows by
\begin{eqnarray*}
&& M\ket{\psi^{0,0}_{i_1, \ldots, i_j}} = \frac{1}{\sqrt{n-j-1 \choose t-1-j}} 
\mathop{\sum_{x:|x|=t-1, x_1=0}}_{x_{i_1}=\cdots=x_{i_j}=1,} M \ket{x}  \\
&= &\frac{1}{\sqrt{n-j-1 \choose t-1-j}} 
\mathop{\sum_{x:|x|=t-1, x_1=0}}_{x_{i_1}=\cdots=x_{i_j}=1} \sum_{\ell:x_\ell=1} 
\ket{1 x_2 \ldots x_{\ell-1} 0 x_{\ell+1} \ldots x_n} \\
&= &\frac{n-t+1}{\sqrt{n-j-1 \choose t-1-j}}
\mathop{\sum_{y:|y|=t-1, y_1=1}}_{y_{i_1}=\cdots=y_{i_j}=1} \ket{y} \\
&&+ \frac{1}{\sqrt{n-j-1 \choose t-1-j}}
\sum_{\ell=1}^j \mathop{\sum_{y:|y|=t-1, y_1=1, y_{i_\ell}=0}}_{y_{i_1}=\cdots=y_{i_j}=1} \ket{y} \\
&= &\frac{n-t-j+1}{\sqrt{n-j-1 \choose t-1-j}}
\mathop{\sum_{y:|y|=t-1, y_1=1}}_{y_{i_1}=\cdots=y_{i_j}=1} \ket{y} \\
&&+ \frac{1}{\sqrt{n-j-1 \choose t-1-j}}
\sum_{\ell=1}^j \mathop{\mathop{\sum_{y:|y|=t-1, y_1=1}}_{y_{i_1}=\cdots=y_{i_{\ell-1}}=1}}_{
y_{i_{\ell+1}}=\cdots=y_{i_j}=1} \ket{y} \\
&= &(n-t-j+1)\sqrt{\frac {t-1-j} {n-t+1}} \ket{\psi^{0,1}_{i_1, \ldots, i_j}}\\ 
&& + \sqrt{\frac{n-j}{n-t+1}} \sum_{\ell=1}^j
 \ket{\psi^{0,1}_{i_1, \ldots, i_{\ell-1}, i_{\ell+1}, \ldots, i_j}}.
\end{eqnarray*}
This proves (\ref{eq-rem4}), with $\ket{\psi'}$ equal to the second term.

Second, for every $j$, $M(T_{j,0,0})\subseteq T_{j,0,1}$ and 
$M (T_{j,0,0}^{\perp}) \subseteq T_{j,0,1}^{\perp}$.
The first statement follows from equation (\ref{eq-rem4}),
because the subspaces $T_{j,0,0}$, $T_{j,0,1}$ are spanned by the states 
$\ket{\psi^{0,0}_{i_1, \ldots, i_j}}$ and $\ket{\psi^{0,1}_{i_1, \ldots, i_j}}$,
respectively, and $T_{j-1,0,1} \subseteq T_{j,0,1}$.
To prove the second statement, let $\ket{\psi}\in T_{j,0,0}^{\perp}$, 
$\ket{\psi}=\sum_{x} a_{x}\ket{x}$. 
We would like to prove $M\ket{\psi}\in T_{j,0,1}^{\perp}$.
This is equivalent to $\bra{\psi^{0,1}_{i_1, \ldots, i_j}}M\ket{\psi}=0$
for all $i_1, \ldots, i_j$. 
We have 
\begin{align*}
  \bra{\psi^{0,1}_{i_1, \ldots, i_j}}M\ket{\psi} &= \frac{1}{\sqrt{n-j-1
\choose t-j-2}} 
\mathop{\sum_{y:|y|=t-1, y_1=1}}_{y_{i_1}=\cdots=y_{i_j}=1} \bra{y}M\ket{\psi} \\
&=\frac{1}{\sqrt{n-j-1 \choose t-j-2}} 
\mathop{\sum_{x:|x|=t-1, x_1=0}}_{x_{i_1}=\cdots=x_{i_j}=1}
\mathop{\sum_{\ell:x_\ell=1}}_{\ell\notin \{i_1, \ldots, i_j\}} a_x  \\
&= \frac{t-1-j}{\sqrt{n-j-1 \choose t-j-2}} 
\mathop{\sum_{x:|x|=t-1, x_1=0}}_{x_{i_1}=\cdots=x_{i_j}=1} a_x = 0.
\end{align*}
The first equality follows by writing out $\bra{\psi^{0,1}_{i_1, \ldots, i_j}}$,
the second equality follows by writing out $M$. The third equality follows because,
for every $x$ with $|x|=t-1$ and $x_{i_1}=\cdots=x_{i_j}=1$, 
there are $t-1-j$ more $\ell\in[n]$ satisfying $x_\ell=1$.
The fourth equality follows because $\sum_{x:|x|=t-1, x_1=0 \atop x_{i_1}=\cdots=x_{i_j}=1} a_x$ is a constant
times $\bra{\psi^{0,0}_{i_1, \ldots, i_j}}\psi\rket$, and 
$\bra{\psi^{0,0}_{i_1, \ldots, i_j}}\psi\rket=0$ because $\ket{\psi}\in
T_{j,0,0}^{\perp}$.  

To deduce equation \eqnref{eq:map}, we write 
\[\ket{\psi^{0,0}_{i_1, \ldots, i_j}}=\ket{\tilde{\psi}^{0,0}_{i_1, \ldots, i_j}}+
\P_{T_{j-1, 0, 0}} \ket{\psi^{0,0}_{i_1, \ldots, i_j}} .\]
Since $M(T_{j-1, 0, 0})\subseteq T_{j-1, 0, 1}$ and 
$M(T_{j-1, 0, 0}^{\perp})\subseteq T_{j-1, 0, 1}^{\perp}$, 
\begin{align*} 
M \ket{\tilde{\psi}^{0,0}_{i_1, \ldots, i_j}} &= \P_{T_{j-1, 0, 1}^{\perp}} M 
\ket{\psi^{0,0}_{i_1, \ldots, i_j}} \\
&= c \P_{T_{j-1, 0, 1}^{\perp}} \ket{\psi^{0, 1}_{i_1, \ldots, i_j}} 
= c \ket{\tilde{\psi}^{0, 1}_{i_1, \ldots, i_j}} ,
\end{align*}
with the second equality following from (\ref{eq-rem4}) and $\ket{\psi'}\in T_{j-1, 0, 1}$.
This proves the first half of \eqnref{eq:map}. The second half follows similarly.
Therefore
\[ 
 \bra{\tilde{\psi}^{0, 0}_{i_1, \ldots, i_j}} 
M^{\dagger} M 
\ket{\tilde{\psi}^{0, 0}_{i'_1, \ldots, i'_j}} =
 c\cdot c' \bra{\tilde{\psi}^{0, 0}_{i_1, \ldots, i_j}}
\tilde{\psi}^{0, 0}_{i'_1, \ldots, i'_j}\rket .\]
Hence $M$ is a multiple of a unitary transformation. 
By equation \eqnref{eq:map}, $U_{01}=M/c$ and, therefore,
$U_{01}$ is also a multiple of a unitary transformation.

Next, we define $M$ by
$M\ket{0 x_2 \ldots x_n}=\ket{1 x_2 \ldots x_n}$.
Then $M$ is a unitary transformation from the space spanned by
$\ket{0 x_2 \ldots x_n}$, $x_2+\cdots+x_2=t-1$, to
the space spanned by $\ket{1 x_2 \ldots x_n}$, $1+x_2+\cdots+x_n=t$.
We claim that $U_{11}=M$.
To prove that, we first observe that 
\begin{align*}
M & \ket{\psi^{0, 0}_{i_1, \ldots, i_j}} 
= \frac{1}{\sqrt{{n-j-1 \choose t-j-1}}} \mathop{\sum_{x_2, \ldots, x_n:}}_{x_{i_1}=\cdots=x_{i_j}=1} M \ket{0 x_2 \ldots x_n}\\ 
& = \frac{1}{\sqrt{{n-j-1 \choose t-j-1}}} \mathop{\sum_{x_2, \ldots, x_n:}}_{x_{i_1}=\cdots=x_{i_j}=1} \ket{1 x_2 \ldots x_n} 
 = \ket{\psi^{1, 1}_{i_1, \ldots, i_j}}. 
\end{align*}
Since $T_{j, a, b}$ is defined as the subspace spanned by all
$\ket{\psi^{a, b}_{i_1, \ldots, i_j}}$, this means that
$M (T_{j, 0, 0})= T_{j, 1, 1}$ and similarly   
$M (T_{j-1, 0, 0})= T_{j-1, 1, 1}$.
Since $M$ is unitary, this implies $M (T_{j-1, 0, 0}^{\perp})= 
T_{j-1, 1, 1}^{\perp}$ and
\begin{align*}
M \ket{\tilde{\psi}^{0, 0}_{i_1, \ldots, i_j}} &=
M \P_{T_{j-1, 0, 0}^{\perp}} 
\ket{\psi^{0, 0}_{i_1, \ldots, i_j}}\\
 &= \P_{T_{j-1, 1, 1}^{\perp}} \ket{\psi^{1, 1}_{i_1, \ldots, i_j}} =
\ket{\tilde{\psi}^{1, 1}_{i_1, \ldots, i_j}} .
\end{align*}

Finally, we have $U_{10}=U''_{10} U_{11}$, where $U''_{10}$ is defined by
$U''_{10}\ket{\tilde{\psi}^{1, 1}_{i_1, \ldots, i_j}}=
\ket{\tilde{\psi}^{1, 0}_{i_1, \ldots, i_j}}$.
Since $U_{11}$ is unitary, 
it suffices to prove that $U''_{10}$ is a multiple of a unitary 
transformation and this follows similarly to $U_{01}$ being
a multiple of a unitary transformation.
\end{proof}

Let $\ket{\psi_{00}}$ be an arbitrary state in $S_{j, 0, 0}$ for some $j\in\{0, \ldots, t-1\}$.
Define $\ket{\psi_{ab}}=U'_{ab}\ket{\psi_{00}}$ for $ab\in\{01, 10, 11\}$.
Let $\ket{\psi_2},\ldots,\ket{\psi_k}$ be vectors
from subspaces $R_{j_2},\ldots,R_{j_k}$, 
for some $j_2, \ldots, j_k$.
We first analyze the case when $\rho_{d, 1}$ belongs to the
subspace $\H_4$ spanned by 
$\ket{\psi_{ab}} \otimes \ket{\psi_2} \otimes \cdots \otimes \ket{\psi_k}.$

\begin{claim}
\label{claim:subspace}
Let\\ $\alpha'_a=\sqrt{\frac{n-(t-1+a)}{n-j}} \|\tilde{\psi}^{a, 0}_{i_1, \ldots, i_j}\|$,
$\beta'_a=\sqrt{\frac{(t-1+a)-j}{n-j}} \|\tilde{\psi}^{a, 1}_{i_1, \ldots, i_j}\|$,\\
$\alpha_a=\frac{\alpha'_a}{\sqrt{(\alpha'_a)^2+(\beta'_a)^2}}$,
$\beta_a=\frac{\beta'_a}{\sqrt{(\alpha'_a)^2+(\beta'_a)^2}}$.
Then 
\begin{enumerate}
\item
$\ket{\phi_1}=\alpha_0\ket{\psi_{00}}+\beta_0 \ket{\psi_{01}}+
\alpha_1\ket{\psi_{10}}+\beta_1 \ket{\psi_{11}}$ belongs to $S_{j, +}$;
\item
$\ket{\phi_2}=\beta_0\ket{\psi_{00}}-\alpha_0 \ket{\psi_{01}}+
\beta_1\ket{\psi_{10}}-\alpha_1 \ket{\psi_{11}}$ belongs to $S_{j+1, +}$;
\item
Any linear combination of $\ket{\psi_{00}}$, $\ket{\psi_{01}}$,
$\ket{\psi_{10}}$ and $\ket{\psi_{11}}$ which is orthogonal to $\ket{\phi_1}$
and $\ket{\phi_2}$ belongs to $S_{-} = \bigoplus_{j=0}^{t} S_{j,-}$.
\end{enumerate}
\end{claim}

\begin{proof}
Let $i_1, \ldots, i_j$ be $j$ distinct elements of $\{2, \ldots, n\}$.
As shown in the beginning of the proof of \clmref{clm:relate}, 
\begin{align*}
\ket{\tilde{\psi}^a_{i_1, \ldots, i_j}} &=
\sqrt{\frac{n-(t-1+a)}{n-j}} \ket{\tilde{\psi}^{a, 0}_{i_1, \ldots, i_j}}\\
& + \sqrt{\frac{(t-1+a)-j}{n-j}} \ket{\tilde{\psi}^{a, 1}_{i_1, \ldots, i_j}}\\ 
& = \alpha'_a \frac{\ket{\tilde{\psi}^{a, 0}_{i_1, \ldots, i_j}}}{
\| \tilde{\psi}^{a, 0}_{i_1, \ldots, i_j}\|}
+ \beta'_a
\frac{\ket{\tilde{\psi}^{a, 1}_{i_1, \ldots, i_j}}}{
\| \tilde{\psi}^{a, 1}_{i_1, \ldots, i_j}\|}.
\end{align*}
This means that  $\|\tilde{\psi}^a_{i_1, \ldots, i_j}\|=
\sqrt{(\alpha'_a)^2+(\beta'_a)^2}$ and 
\[ \frac{\ket{\tilde{\psi}^a_{i_1, \ldots, i_j}}}{
\|\tilde{\psi}^a_{i_1, \ldots, i_j}\|} = 
\alpha_a \frac{\ket{\tilde{\psi}^{a, 0}_{i_1, \ldots, i_j}}}{
\| \tilde{\psi}^{a, 0}_{i_1, \ldots, i_j}\|}
+ \beta_a
\frac{\ket{\tilde{\psi}^{a, 1}_{i_1, \ldots, i_j}}}{
\| \tilde{\psi}^{a, 1}_{i_1, \ldots, i_j}\|} .\]
Since the states $\ket{\tilde{\psi}^0_{i_1, \ldots, i_j}}$
span $S_{j, 0}$, the state $\ket{\psi_{00}}$
is a linear combination of states 
$\frac{\ket{\tilde{\psi}^{0,0}_{i_1, \ldots, i_j}}}{
\|\tilde{\psi}^{0,0}_{i_1, \ldots, i_j}\|}$.
By \clmref{claim:unitary}, the states $\ket{\psi_{ab}}$
are linear combinations of
$\frac{\ket{\tilde{\psi}^{a,b}_{i_1, \ldots, i_j}}}{
\|\tilde{\psi}^{a,b}_{i_1, \ldots, i_j}\|}$ with the same
coefficients. Therefore, $\ket{\phi_1}$ is a
linear combination of
\[ \alpha_0 \frac{\ket{\tilde{\psi}^{0,0}_{i_1, \ldots, i_j}}}{
\|\tilde{\psi}^{0,0}_{i_1, \ldots, i_j}\|}+
\beta_0 \frac{\ket{\tilde{\psi}^{0,1}_{i_1, \ldots, i_j}}}{
\|\tilde{\psi}^{0,1}_{i_1, \ldots, i_j}\|}+
\alpha_1 \frac{\ket{\tilde{\psi}^{1,0}_{i_1, \ldots, i_j}}}{
\|\tilde{\psi}^{1,0}_{i_1, \ldots, i_j}\|}+
\beta_1 \frac{\ket{\tilde{\psi}^{1,1}_{i_1, \ldots, i_j}}}{
\|\tilde{\psi}^{1,1}_{i_1, \ldots, i_j}\|}\]
\[ =
\frac{\ket{\tilde{\psi}^0_{i_1, \ldots, i_j}}}{
\|\tilde{\psi}^0_{i_1, \ldots, i_j}\|} + 
\frac{\ket{\tilde{\psi}^1_{i_1, \ldots, i_j}}}{
\|\tilde{\psi}^1_{i_1, \ldots, i_j}\|} ,\]
each of which, by definition, belongs to $S_{j, +}$.

Let $i_1, \ldots, i_{j}$ be distinct elements of 
$\{2, \ldots, n\}$.
We claim 
\begin{equation}
\label{eq:phi2} 
\frac{\ket{\tilde{\psi}^a_{1, i_1, \ldots, i_j}}}{
\| \tilde{\psi}^a_{1, i_1, \ldots, i_j}\|} = 
\beta_a \frac{\ket{\tilde{\psi}^{a, 0}_{i_1, \ldots, i_j}}}{
\| \tilde{\psi}^{a, 0}_{i_1, \ldots, i_j}\|}
- \alpha_a
\frac{\ket{\tilde{\psi}^{a, 1}_{i_1, \ldots, i_j}}}{
\| \tilde{\psi}^{a, 1}_{i_1, \ldots, i_j}\|} .
\end{equation}
By \clmref{clm:relate}, the right hand side of \eqnref{eq:phi2}
belongs to $S_{j+1, a}$. We need to show that it is equal to
$\ket{\tilde{\psi}^a_{1, i_1, \ldots, i_j}}$.
We have
\begin{align*}
\ket{\tilde{\psi}^a_{1, i_1, \ldots, i_j}} &=
\P_{T_{j, a}^{\perp}} \ket{\psi^a_{1, i_1, \ldots, i_j}} =
\P_{T_{j, a}^{\perp}} \ket{\psi^{a, 1}_{i_1, \ldots, i_j}}
\\
&= \P_{T_{j, a}^{\perp}} \P_{T_{j-1, a, 1}^{\perp}}
\ket{\psi^{a, 1}_{i_1, \ldots, i_j}} =
\P_{T_{j, a}^{\perp}} \ket{\tilde{\psi}^{a, 1}_{i_1, \ldots, i_j}} ,
\end{align*}
where the third equality follows from $T_{j-1, a, 1}\subseteq T_{j, a}$.
This is because the states $\ket{\psi^{a, 1}_{i_1, \ldots, i_{j-1}}}$ spanning
$T_{j-1,a,1}$ are the same as the states $\ket{\psi^a_{1, i_1, \ldots,
i_{j-1}}}$ in $T_{j,a}$.
Write
\[ \ket{\tilde{\psi}^{a, 1}_{i_1, \ldots, i_j}}=c_1\ket{\delta_1}+c_2\ket{\delta_2}\]
where 
\begin{align*}
\ket{\delta_1} & =\alpha_a
\frac{\ket{\tilde{\psi}^{a, 0}_{i_1, \ldots, i_j}}}{
\|\tilde{\psi}^{a, 0}_{i_1, \ldots, i_j}\|}
+\beta_a\frac{\ket{\tilde{\psi}^{a, 1}_{i_1, \ldots, i_j}}}{
\|\tilde{\psi}^{a, 1}_{i_1, \ldots, i_j}\|},\\
\ket{\delta_2} &=\beta_a\frac{\ket{\tilde{\psi}^{a, 0}_{i_1, \ldots, i_j}}}{
\|\tilde{\psi}^{a, 0}_{i_1, \ldots, i_j}\|}
-\alpha_a\frac{\ket{\tilde{\psi}^{a, 1}_{i_1, \ldots, i_j}}}{
\|\tilde{\psi}^{a, 1}_{i_1, \ldots, i_j}\|} .
\end{align*}
By \clmref{clm:relate}, we have $\ket{\delta_1}\in S_{j, a}\subseteq T_{j, a}$, 
$\ket{\delta_2}\in S_{j+1, a}\subseteq T_{j, a}^{\perp}$.
Therefore, $\P_{T_{j, a}^{\perp}} \ket{\tilde{\psi}^{a, 1}_{i_1, \ldots, i_j}}=c_2\ket{\delta_2}$ and
\[ \frac{\ket{\tilde{\psi}^a_{1, i_1, \ldots, i_j}}}{\|\tilde{\psi}^a_{1, i_1, \ldots, i_j}\|}=
\ket{\delta_2} = \beta_a \frac{\ket{\tilde{\psi}^{a, 0}_{i_1, \ldots, i_j}}}{
\| \tilde{\psi^{a, 0}_{i_1, \ldots, i_j}}\|}
- \alpha_a \frac{\ket{\tilde{\psi}^{a, 1}_{i_1, \ldots, i_j}}}{
\| \tilde{\psi}^{a, 1}_{i_1, \ldots, i_j}\|} ,\]
proving \eqnref{eq:phi2}.

Similarly to the argument for $\ket{\phi_1}$, equation \eqnref{eq:phi2} implies that
$\ket{\phi_2}$ is a
linear combination of
\[ \beta_0 \frac{\ket{\tilde{\psi}^{0,0}_{i_1, \ldots, i_j}}}{
\|\tilde{\psi}^{0,0}_{i_1, \ldots, i_j}\|}-
\alpha_0 \frac{\ket{\tilde{\psi}^{0,1}_{i_1, \ldots, i_j}}}{
\|\tilde{\psi}^{0,1}_{i_1, \ldots, i_j}\|}+
\beta_1 \frac{\ket{\tilde{\psi}^{1,0}_{i_1, \ldots, i_j}}}{
\|\tilde{\psi}^{1,0}_{i_1, \ldots, i_j}\|}-
\alpha_1 \frac{\ket{\tilde{\psi}^{1,1}_{i_1, \ldots, i_j}}}{
\|\tilde{\psi}^{1,1}_{i_1, \ldots, i_j}\|}
\]
\[
= \frac{\ket{\tilde{\psi}^0_{1, i_1, \ldots, i_j}}}{
\|\tilde{\psi}^0_{1, i_1, \ldots, i_j}\|} + 
\frac{\ket{\tilde{\psi}^1_{1, i_1, \ldots, i_j}}}{
\|\tilde{\psi}^1_{1, i_1, \ldots, i_j}\|} \]
and each of those states belongs to $S_{j+1, +}$.

To prove the third part of \clmref{claim:subspace}, we observe that any
vector orthogonal to $\ket{\phi_{1}}$ and $\ket{\phi_2}$ is a linear combination
of 
\[ \ket{\phi_3} = \alpha_0 \ket{\psi_{00}}+\beta_0 \ket{\psi_{01}} -
\alpha_1 \ket{\psi_{10}}-\beta_1 \ket{\psi_{11}},\]
which, in turn, is a linear combination of vectors
\[ \frac{\ket{\tilde{\psi}^0_{i_1, \ldots, i_j}}}{
\|\tilde{\psi}^0_{i_1, \ldots, i_j}\|} - 
\frac{\ket{\tilde{\psi}^1_{i_1, \ldots, i_j}}}{
\|\tilde{\psi}^1_{i_1, \ldots, i_j}\|} \]
and
\[ \ket{\phi_4} = \beta_0 \ket{\psi_{00}}-\alpha_0 \ket{\psi_{01}} -
\beta_1 \ket{\psi_{10}}+\alpha_1 \ket{\psi_{11}}\]
which is a linear combination of vectors
\[ \frac{\ket{\tilde{\psi}^0_{1, i_1, \ldots, i_j}}}{
\|\tilde{\psi}^0_{1, i_1, \ldots, i_j}\|} - 
\frac{\ket{\tilde{\psi}^1_{1, i_1, \ldots, i_j}}}{
\|\tilde{\psi}^1_{1,i_1, \ldots, i_j}\|} .
\]
This means that we have $\ket{\phi_3}\in S_{j, -}$ and $\ket{\phi_4}\in 
S_{j+1, -}$.
\end{proof}

\vfill\eject 

\begin{claim}
\label{claim:norm}
Let $j < t/2$ and $x^\ul j = x (x-1) \cdots (x-j+1)$.
\begin{enumerate}
\item $\displaystyle\|\tilde\psi^{a,b}_{i_1,\ldots,i_j}\| = \sqrt{ \frac {(n-t-a+b)^{\ul
j}} {(n-j)^{\ul j}} }$.
\item $\displaystyle\|\tilde\psi^{a,0}_{i_1,\ldots,i_j}\| \ge \frac 1 {\sqrt 2}
\|\tilde\psi^{a,1}_{i_1,\ldots,i_j}\|$.
\item $\displaystyle\frac {\|\tilde\psi^{0,0}_{i_1,\ldots,i_j}\| \cdot
\|\tilde\psi^{1,1}_{i_1,\ldots,i_j}\|} {\|\tilde\psi^{0,1}_{i_1,\ldots,i_j}\|
\cdot \|\tilde\psi^{1,0}_{i_1,\ldots,i_j}\|} = 1 + O\left(\frac 1 t\right)$.
\end{enumerate}
\end{claim}

\begin{proof}
Define $t_a = t-1+a$.
We calculate the vector
\[ \ket{\tilde{\psi}^{a, b}_{i_1, \ldots, i_j}}=\P_{T_{j-1,a,b}^{\perp}} 
\ket{\psi^{a, b}_{i_1, \ldots, i_j}} .\]
Both vector $\ket{\psi^{a, b}_{i_1, \ldots, i_j}}$ and 
subspace $T_{j-1,a,b}$ are fixed by 
\[ U_\pi \ket{x}=\ket{x_{\pi(1)}\ldots x_{\pi(n)}} \]
for any permutation $\pi$ that fixes $1$ and
maps $\{i_1, \ldots, i_j\}$ to itself.
Hence $\ket{\tilde{\psi}^{a, b}_{i_1, \ldots, i_j}}$
is fixed by any such $U_{\pi}$ as well.
Therefore, the amplitude of $\ket{x}$ with $|x|=t_a$, $x_1=b$ in 
$\ket{\tilde{\psi}^{a, b}_{i_1, \ldots, i_j}}$ only depends on
$|\{i_1, \ldots, i_j\}\cap \{i:x_i=1\}|$, so
$\ket{\tilde{\psi}^{a, b}_{i_1, \ldots, i_j}}$ is of the form
\[ \ket{\upsilon_{a,b}}=\sum_{m=0}^j \kappa_m \mathop{\sum_{x:|x|=t_a, 
x_1=b}}_{
|\{i_1, \ldots, i_j\}\cap \{i:x_i=1\}| = m} \ket{x} .\]
To simplify the following calculations, we multiply 
$\kappa_0, \ldots, \kappa_j$ by the same constant 
so that $\kappa_j=1/\sqrt{{n-j-1 \choose t_a-j-b}}$. 
Then 
$\ket{\tilde{\psi}^{a, b}_{i_1, \ldots, i_j}}$
remains a multiple of $\ket{\upsilon_{a,b}}$ but may no longer be
equal to $\ket{\upsilon_{a,b}}$. 

$\kappa_0, \ldots, \kappa_{j-1}$ should be such that the state
is orthogonal to $T_{j-1,a,b}$ and, in particular, orthogonal 
to the states $\ket{\psi^{a, b}_{i_1, \ldots, i_\ell}}$ for all
$\ell\in\{0, \ldots, j-1\}$. By writing out
$\lbra \upsilon_{a,b} \ket{\psi^{a, b}_{i_1, \ldots, i_\ell}}=0$:
\begin{equation}
\label{eq-system} 
\sum_{m=\ell}^j \kappa_m {n-j-1 \choose t_a-m-b} {j-\ell \choose m-\ell} 
=0   .
\end{equation}
To show that, we first note that
$\ket{\psi^{a, b}_{i_1, \ldots, i_\ell}}$ is a uniform superposition
of all $\ket{x}$ with $|x|=t_a$, $x_1=b$, $x_{i_1}=\cdots=x_{i_\ell}=1$.
If we want to choose $x$ subject to those constraints
and also satisfying $|\{i_1, \ldots, i_j\}\cap \{i:x_i=1\}|=m$,
then we have to set $x_i=1$ for $m-\ell$ different $i\in\{i_{\ell+1}, \ldots, 
i_j\}$
and for $t_a-m-b$ different $i\notin\{1, i_1, \ldots, i_j\}$. 
This can be done in ${j-\ell \choose m-\ell}$
and ${n-j-1 \choose t_a-m-b}$ different ways, respectively.

By solving the system of equations (\ref{eq-system}),
starting from $\ell=j-1$ and going down to $\ell=0$,
we get
that the only solution is
\begin{equation}
\label{eq-solution} 
 \kappa_m=(-1)^{j-m}
\frac{{n-j-1 \choose t_a-j-b}}{{n-j-1 \choose t_a-m-b}} \kappa_j 
.\end{equation}

Let $\ket{\upsilon'_{a,b}}=\frac{\ket{\upsilon_{a,b}}}{\|\upsilon_{a,b}\|}$ be the normalized
version of $\ket{\upsilon_{a,b}}$. Then
\begin{align}
\ket{\tilde{\psi}^{a, b}_{i_1, \ldots, i_j}} &= 
\lbra \upsilon'_{a,b}\ket{\psi^{a, b}_{i_1, \ldots, i_j}} \ket{\upsilon'_{a,b}} ,
\nonumber \\
\label{eq-2807} 
\| \tilde{\psi}^{a, b}_{i_1, \ldots, i_j} \| &= 
\lbra \upsilon'_{a,b}\ket{\psi^{a, b}_{i_1, \ldots, i_j}} = 
\frac{\lbra \upsilon_{a,b}\ket{\psi^{a, b}_{i_1, \ldots, i_j}}}{\|\upsilon_{a,b}\|} .
\end{align}
We have 
\[ \lbra \upsilon_{a,b} \ket{\psi^{a, b}_{i_1, \ldots, i_j}} = 1 ,\]
because $\ket{\psi^{a, b}_{i_1, \ldots, i_j}}$ consists of 
${n-j-1 \choose t_a-j-b}$
basis states $\ket{x}$, $x_1=b$, $x_{i_1}=\cdots=x_{i_j}=1$, each 
having amplitude $1/\sqrt{{n-j-1 \choose t_a-j-b}}$
in both $\ket{\upsilon_{a,b}}$ and $\ket{\psi^{a, b}_{i_1, \ldots, i_j}}$.
Furthermore,
\begin{align}
\|\upsilon_{a,b} \|^2 &= \sum_{m=0}^j {j \choose m} {n-j-1 \choose 
t_a-m-b} \kappa_m^2
\nonumber \\
&=
\sum_{m=0}^j {j \choose m} \frac{{n-j-1 \choose t_a-j-b}^2}{
{n-j-1 \choose t_a-m-b}} \kappa_j^2\nonumber\\
& =
\sum_{m=0}^j {j \choose m} \frac{{n-j-1 \choose t_a-j-b}}{{n-j-1 \choose 
t_a-m-b}}
\nonumber \\
&=
\sum_{m=0}^j {j \choose m} 
\frac{(t_a-m-b)!(n-t_a+m-j-1+b)!}{(t_a-j-b)!(n-t_a-1+b)!}
\nonumber \\
\label{eq-2807a} 
&= \sum_{m=0}^j {j \choose m} \frac{(t_a-m-b)^\ul{j-m}} {(n-t_a-1+b)^\ul{j-m}} .
\end{align}
Here the first equality follows because there are 
${j \choose m}{n-j-1\choose t_a-m-b}$ vectors $x$ such that
$|x|=t_a$, $x_1=b$, $x_i=1$ for $m$ different $i\in\{i_1, \ldots, i_j\}$
and $t_a-m$ different $i\notin\{1, i_1, \ldots, i_j\}$,
the second equality follows from equation 
(\ref{eq-solution}) and the third equality follows from our choice 
$\kappa_j=1/\sqrt{{n-j-1 \choose t_a-j-b}}$. 

{}From equations \eqnref{eq-2807} and \eqnref{eq-2807a},
$\|\tilde{\psi}^{a, b}_{i_1, \ldots, i_j} \| 
= \frac{1}{\sqrt{A_{a, b}}}$ where 
$A_{a,b} = \sum_{m=0}^\infty C_{a,b}(m)$ and
\[
C_{a,b}(m) = {j \choose m} \frac {(t_a-m-b)^\ul{j-m}}
  {(n-t_a-1+b)^\ul{j-m}}.
\]
The terms with $m > j$ are zero because ${j \choose m} = 0$ for $m > j$.

We compute the combinatorial sum $A_{a,b}$ using hypergeometric series
\cite[Section 5.5]{gkp:concrete}.
Since
\[
\frac {C_{a,b}(m+1)} {C_{a,b}(m)} =
  \frac{ (m-j) (m+n-t_a-j+b) } { (m+1) (m-t_a+b) }
\]
is a rational function of $m$, $A_{a,b}$ is a hypergeometric series and its value is
\[
A_{a,b} = \sum_{m=0}^\infty C_{a,b}(m)
  = C_{a,b}(0) \cdot F\Big({-j,\ n-t_a-j+b \atop -t_a+b} \Big\vert 1\Big).
\]
We apply Vandermonde's convolution $F({-j,\ x \atop y} \vert 1) = (x-y)^\ul j / (-y)^\ul j$
\cite[Equation 5.93 on page 212]{gkp:concrete}, which holds for every
integer $j \ge 0$, and obtain
\[
A_{a,b} = \frac {(t_a-b)^\ul j} {(n-t_a-1+b)^\ul j} \cdot
  \frac {(n-j)^\ul j} {(t_a-b)^\ul j}
  = \frac {(n-j)^\ul j} {(n-t_a-1+b)^\ul j}.
\]
This proves the first part of the claim, that
$\|\tilde\psi^{a,b}_{i_1,\ldots,i_j}\| = \sqrt{ (n-t_a-1+b)^{\ul j} / (n-j)^{\ul
j} }$. 

The second part of the claim follows because
\begin{align*}
\frac {\|\tilde\psi^{a,0}_{i_1,\ldots,i_j}\|} 
  {\|\tilde\psi^{a,1}_{i_1,\ldots,i_j}\|}
&= \sqrt{ \frac {(n-t_a-1)^{\ul j}} {(n-t_a)^{\ul j}} }
= \sqrt{ \frac {n-t_a-j} {n-t_a} } \\
&= \sqrt{ 1 - \frac j {n-t_a} }
\ge \sqrt{ 1 - \frac {n/4} {n/2} }
= \frac 1 {\sqrt 2} ,
\end{align*}
because $j\leq t_a/2$, and $t_a \le n/2$.

For the third part,
\begin{align*}
\frac {A_{1,0} A_{0,1}} {A_{0,0} A_{1,1}} 
  & = \frac {((n-t)^\ul j)^2} {(n-t+1)^\ul j (n-t-1)^\ul j}\\
  & = \frac {(n-t) (n-t-j+1)} {(n-t+1) (n-t-j)}\\
  & = 1 + \frac j {(n-t+1) (n-t-j)},
\end{align*}
which is $1 + \Theta(j/n^2) = 1 + O(1/t)$ for $t \leq n/2$ and $j \leq t/2$.
The expression in the third part of the claim is the square root of this value, hence it is $1 + O(1/t)$.
\end{proof}

\begin{claim}
\label{claim:beta}
If $j< t/2$, then $\beta_a\leq \sqrt{\frac{2 t} n}$.
\end{claim}

\begin{proof}
Define $t_a = t-1+a$.
By \clmref{claim:norm},
$\|\tilde{\psi}^{a, 0}_{i_1, \ldots, i_j}\| 
\geq \frac{1}{\sqrt{2}} \|\tilde{\psi}^{a, 1}_{i_1, \ldots, i_j}\|$.
That implies
\begin{align*}
  \alpha'_a &= \frac{\sqrt{n-t_a}}{\sqrt{n-j}} 
\|\tilde{\psi}^{a, 0}_{i_1, \ldots, i_j}\| \\
&\geq  \frac{1}{\sqrt{2}} 
\frac{\sqrt{n-t_a}}{\sqrt{t_a-j}} \frac{\sqrt{t_a-j}}{\sqrt{n-j}}
\|\tilde{\psi}^{a, 1}_{i_1, \ldots, i_j}\|
= \frac{\sqrt{n-t_a}}{\sqrt{2(t_a-j)}} \beta'_a
\end{align*}
and hence
\[
\sqrt{(\alpha'_a)^2+(\beta'_a)^2} \geq
\beta'_a \sqrt{\frac{n-t_a}{2(t_a-j)} + 1} = 
\beta'_a \frac{\sqrt{n+ t_a-2j}}{\sqrt{2(t_a-j)}} .
\]
Then, using $j \leq \frac {t_a} 2$,\\
\hspace*{2em}
$
\displaystyle\beta_a =
\frac{\beta'_a}{\sqrt{(\alpha'_a)^2+(\beta'_a)^2}} \leq
\frac{\sqrt{2(t_a-j)}}{\sqrt{n+ t_a-2j}} \leq
\sqrt{\frac {2 t} n}.
$
\end{proof}

\begin{claim}
\label{claim:beta1}
If $j< t/2$, then $\displaystyle |\alpha_0\beta_1-\alpha_1\beta_0| = O \left(\frac{1}{\sqrt{t n}} \right).$ 
\end{claim}

\begin{proof}
We first estimate 
\[ \frac{\alpha_0 \beta_1}{\alpha_1 \beta_0} =
\frac{\alpha'_0 \beta'_1}{\alpha'_1 \beta'_0} = 
\frac{\sqrt{(n-t+1) (t-j)}}{\sqrt{(n-t)(t-1-j)}} \cdot
\frac{\|\tilde{\psi}^{0, 0}_{i_1, \ldots, i_j}\| 
\|\tilde{\psi}^{1, 1}_{i_1, \ldots, i_j}\|}{
\|\tilde{\psi}^{1, 0}_{i_1, \ldots, i_j}\| 
\|\tilde{\psi}^{0, 1}_{i_1, \ldots, i_j}\|} .\]

By \clmref{claim:norm}, we have 
\[ \frac{\alpha'_0 \beta'_1}{\alpha'_1 \beta'_0} = 
\left( 1+ O\left(\frac{1}{t} \right) \right) 
\frac{\sqrt{(n-t+1) (t-j)}}{\sqrt{(n-t)(t-1-j)}} .\]
Since $\frac{\sqrt{t-j}}{\sqrt{t-1-j}}=\sqrt{1+\frac{1}{t-1-j}}=
1+O(\frac{1}{t-1-j})=1+O(\frac{1}{t})$ and, similarly,
$\frac{\sqrt{n-t+1}}{\sqrt{n-t}}=1+O(\frac{1}{n-t})=1+O(\frac{1}{t})$,
we have shown that $\frac{\alpha_0 \beta_1}{\alpha_1 \beta_0}$ is
of order $1 + O(\frac{1}{t})$.
We thus have
\[
|\alpha_0\beta_1-\beta_0\alpha_1| = 
O\left(\frac{1}{t}\right) |\beta_0\alpha_1|
= O\left(\frac{1}{t} \cdot \sqrt{ \frac t n } \right) 
= O\left(\frac{1}{\sqrt{t n}} \right),
\]
thanks to \clmref{claim:beta} and the fact that $|\alpha_1| \le 1$.
\end{proof}

We pick an orthonormal basis for $\H_4$ that has $\ket{\phi_1}$ and $\ket{\phi_2}$
as its first two vectors. Let $\ket{\phi_3}$ and $\ket{\phi_4}$ be the other
two basis vectors. We define 
\begin{equation}
\label{eq:4basis} 
\ket{\chi_i}=\ket{\phi_i} \otimes \ket{\psi_2} \otimes \cdots \otimes 
\ket{\psi_k}.
\end{equation}
By \clmref{claim:subspace}, $\ket{\chi_1}$ belongs to 
$S_{j, +}\otimes R_{j_2} \otimes \cdots \otimes R_{j_k}$ which is contained in
$\R_{\min(j, t/2)+j_2+\cdots+j_k}$. 
Similarly, $\ket{\chi_2}$ belongs to 
$\R_{\min(j+1, t/2)+j_2+\cdots+j_k}$ and 
$\ket{\chi_3}$, $\ket{\chi_4}$ belong to
$\R_{t/2+j_2+\cdots+j_k}$.
If $j<t/2$, this means that 
\begin{align}
P(\rho_{d, 1}) = &\, q^{j_2+\cdots+j_k}\cdot
\left( q^j \bra{\chi_1}\rho_{d, 1}\ket{\chi_1}+
q^{j+1} \bra{\chi_2}\rho_{d, 1}\ket{\chi_2}\right. \nonumber \\
& \left. + q^{\frac{t}{2}} \bra{\chi_3}\rho_{d, 1}\ket{\chi_3}+
q^{\frac{t}{2}} \bra{\chi_4}\rho_{d, 1}\ket{\chi_4}\right) 
\label{eq:12} 
\end{align}
If $j\geq t/2$, then $\ket{\chi_1}$, $\ket{\chi_2}$,
$\ket{\chi_3}$, $\ket{\chi_4}$ are all in                
$\R_{t/2+j_2+\cdots+j_k}$. This means that
$P(\rho_{d, 1})=q^{t/2+j_2+\cdots+j_k}$
and it remains unchanged by a query.

We define $\gamma_\ell=\bra{\chi_\ell}\rho_{d, 1}\ket{\chi_\ell}$.
Since the support of $\rho_{d, 1}$ is contained in the
subspace spanned by $\ket{\chi_\ell}$, we have
$\gamma_1+\gamma_2+\gamma_3+\gamma_4=\Tr \rho_{d, 1}=1$.
This means that equation \eqnref{eq:12} can be rewritten as
\begin{align}
P(\rho_{d, 1}) &= q^{j+j_2+\cdots+j_k} \gamma_1
+q^{j+j_2+\cdots+j_k+1} \gamma_2 + \nonumber\\
&\hspace{1em} + q^{t/2+j_2+\cdots+j_k} (\gamma_3+\gamma_4) \nonumber \\
&=  \label{eq:12a}
q^{t/2+j_2+\cdots+j_k} + q^{j_2+\cdots+j_k} 
(q^{j+1}-q^{t/2}) (\gamma_1+\gamma_2) + \nonumber \\
&\hspace{1em}  +q^{j_2+\cdots+j_k} (q^j-q^{j+1}) \gamma_1
\end{align}
$P(\rho'_{d, 1})$ can be also expressed in a similar way,
with $\gamma'_j=\bra{\chi_j}\rho'_{d, 1}\ket{\chi_j}$
instead of $\gamma_j$.
By combining equations \eqnref{eq:12a} for $P(\rho_{d, 1})$
and $P(\rho'_{d, 1})$, we get
\begin{align*}
&P(\rho'_{d, 1})-P(\rho_{d, 1}) =
q^{j+j_2+\cdots+j_k} (q^{t/2-j}-q) \\
&\hspace{2em} \cdot(\gamma_1+\gamma_2-\gamma'_1-\gamma'_2)
+q^{j+j_2+\cdots+j_k} (q-1)
(\gamma_1-\gamma'_1).
\end{align*}
Therefore, it suffices to bound 
$|\gamma_1+\gamma_2-\gamma'_1-\gamma'_2|$ and
$|\gamma_1-\gamma'_1|$.
W.l.o.g.~we can assume that $\rho_{d, 1}$ is a pure state 
$\ket{\varphi}\bra{\varphi}$.
Let 
\[ \ket{\varphi}= ( a\ket{\psi_{00}}+b\ket{\psi_{01}}+
c\ket{\psi_{10}}+d\ket{\psi_{11}}
) \otimes \ket{\psi_2} \otimes \dots \otimes \ket{\psi_k}
.\]
Then the state after a query is
\[ \ket{\varphi'}= ( a\ket{\psi_{00}}-b\ket{\psi_{01}}
+c\ket{\psi_{10}}-d\ket{\psi_{11}}
) \otimes \ket{\psi_2} \otimes \dots \otimes \ket{\psi_k}
\]
and we have to bound 
\[ \gamma_\ell-\gamma'_\ell =
|\bra{\chi_\ell}\varphi\rket|^2-|\bra{\chi_\ell}\varphi'\rket|^2 \]
for $\ell\in\{1, 2\}$.
For $\ell=1$, we have 
\[ \bra{\chi_1}\varphi\rket = a \alpha_0+b \beta_0+c 
\alpha_1+ d\beta_1.\]
The expression for $\varphi'$ is similar, with minus signs 
in front of $b\beta_0$ and $d\beta_1$.
Therefore, 
\begin{align}
& \left| |\bra{\chi_1}\varphi\rket|^2 -|\bra{\chi_1}\varphi'\rket|^2 
\right| \nonumber \\
& \hspace{1em} \leq 4|a||b| \alpha_0\beta_0 + 4|c||d| \alpha_1\beta_1 + 4|a| |d| \alpha_0\beta_1+
4|b||c|\alpha_1\beta_0 . 
\label{eq:substract}
\end{align}
Since $|a|$, $|b|$, $|c|$, $|d|$ are all at most $\|\varphi\|=1$ and 
$\alpha_0$, $\alpha_1$ are less than 1, equation \eqnref{eq:substract} 
is at most $8\beta_0+8\beta_1$.
By \clmref{claim:beta}, we have
\[ |\gamma_1 - \gamma'_1| \leq 8\beta_0+8\beta_1
\leq 16\sqrt{\frac{2 t}{n}} .\] 
We also have
\begin{align*}
& \left| \gamma_1 + \gamma_2 - \gamma'_1 -\gamma'_2 \right| \\
& \hspace{1em} = \left| |\bra{\chi_1}\varphi\rket|^2+ |\bra{\chi_2}\varphi\rket|^2 -|\bra{\chi_1}\varphi'\rket|^2 - |\bra{\chi_2}\varphi'\rket|^2\right|  \\
& \hspace{1em} \leq 4|a||d| |\alpha_0 \beta_1 - \alpha_1\beta_0| + 4|b||c| |\alpha_1\beta_0-\alpha_0\beta_1| \\
& \hspace{1em} \leq 8 |\alpha_0 \beta_1 - \alpha_1\beta_0| \leq \frac{8C}{\sqrt{t n}} 
\end{align*}
where $C$ is the big-O constant from \clmref{claim:beta1}.
By taking into account that $P(\rho_{d,1}) \ge q^{j+j_2+\dots+j_k}$,
\begin{align}
P&(\ket{\varphi'}\bra{\varphi'})-P(\ket{\varphi}\bra{\varphi}) \nonumber \\
& \le \left( 
(q^{t/2-j}-q) \frac{8C}{\sqrt{t n}} 
+ (q-1) \frac{16\sqrt{2 t}}{\sqrt{n}} \right) P(\ket{\varphi}\bra{\varphi}) \nonumber \\
& \le \label{eq:4bound} \left( 
(q^{t/2}-1) \frac{8C}{\sqrt{t n}}  
+ (q-1) \frac{16\sqrt{2 t}}{\sqrt{n}} \right) P(\ket{\varphi}\bra{\varphi}). 
\end{align}
This proves \lemref{lem:onestep} for the case when the support of $\rho_{d, 1}$
is contained in $\H_4$. (If $\rho_{d, 1}$ is a mixed state, we just express
it as a mixture of pure states $\ket{\varphi}$. 
The bound for $\rho_{d, 1}$ follows by summing equations \eqnref{eq:4bound}
for every $\ket{\varphi}$.)

For the general case, we divide the entire state space $\H_I$ into 4-dimensional
subspaces. To do that, we first subdivide $\H_I$ into subspaces
\begin{equation}
\label{eq:subdivide} 
(S_{j, 0, 0}\oplus S_{j, 0, 1} \oplus S_{j, 1, 0} 
\oplus S_{j, 1, 1}) \otimes R_{j_2}
\otimes \cdots \otimes R_{j_k} .
\end{equation}
Let states $\ket{\psi^{0, 0}_{1, i}}$, $i\in [\dim S_{j, 0, 0}]$ form a basis for $S_{j, 0, 0}$
and let $\ket{\psi^{a, b}_{1, i}}=U'_{ab} \ket{\psi^{0, 0}_{1, i}}$
for $(a, b)\in\{(0, 1), (1, 0), \penalty0 (1, 1)\}$, where the $U'_{ab}$ 
are the unitaries from~\clmref{claim:unitary}.
Then the $\ket{\psi^{a, b}_{1, i}}$ form a basis for $S_{j, a, b}$.

Let $\ket{\psi_{l, i}}$, $i\in[\dim R_{j_l}]$, form a basis for $R_{j_l}$, 
$l\in\{2, \ldots, k\}$.
We subdivide \eqnref{eq:subdivide} into 4-dimensional
subspaces $H_{i_1, \ldots, i_k}$ spanned by 
\[ \ket{\psi^{a, b}_{1, i_1}} \otimes \ket{\psi_{2, i_2}} \otimes \cdots
\otimes \ket{\psi_{k, i_k}} ,\]
where $a, b$ range over $\{0, 1\}$.
Let $\H_{all}$ be the collection of all $H_{i_1, \ldots, i_k}$
obtained by subdividing all subspaces \eqnref{eq:subdivide}.
We claim that 
\begin{equation}
\label{eq:reduce} 
P(\rho)= \sum_{H \in \H_{all}} P(\P_{H} \rho) .
\end{equation}
Equation~\eqnref{eq:reduce} together with equation \eqnref{eq:4bound}
implies \lemref{lem:onestep}. Since $P(\rho)$ is defined as a weighted sum of 
traces $\Tr \P_{\R_{m}}\rho$, we can prove equation \eqnref{eq:reduce}
by showing
\begin{equation}
\label{eq:trace} 
\Tr \P_{\R_{m}}\rho_{d, 1}= \sum_{H \in \H_{all}} 
\Tr \P_{\R_{m}} \P_{H} \rho_{d, 1} .
\end{equation}
To prove \eqnref{eq:trace}, we define a basis for $\H_I$ by first decomposing
$\H_I$ into subspaces $H\in \H_{all}$, and then for each subspace,
taking the basis consisting of $\ket{\chi_1}$, $\ket{\chi_2}$, 
$\ket{\chi_3}$ and $\ket{\chi_4}$ defined by equation \eqnref{eq:4basis}.
By \clmref{claim:subspace}, each of the basis states belongs 
to one of the subspaces $\R_m$.
This means that each $\R_m$ is spanned by some subset of this basis.

The left hand side of \eqnref{eq:trace} is equal to the sum of squared projections
of $\rho_{d, 1}$ to basis states $\ket{\chi_j}$ that belong to $\R_m$.
Each of the terms $\Tr \P_{\R_{m}} \P_{H} \rho_{d, 1}$ on the right hand
side is equal to the sum of squared projections 
to basis states $\ket{\chi_j}$ that belong to $\R_m\cap H$.
Summing over all $H$ gives the sum of squared projections
of $\rho_{d, 1}$ to all $\ket{\chi_j}$ that belong to $\R_m$.
Therefore, the two sides of \eqnref{eq:trace} are equal.

\balancecolumns

\end{document}